\documentclass[bibyear]{aa} 

\usepackage{natbib}
\setcitestyle{round,aysep={},yysep={;}}

\usepackage{graphicx}
\usepackage{txfonts}
\begin{document}

   \title{Relation of internal attenuation, dust emission, and the size of spiral galaxies}

   \subtitle{Calibration at low-$z$ and how to use it as a cosmological test at high-$z$}

   \author{M. L\'opez-Corredoira\inst{1,2}, C. M. Guti\'errez\inst{1,2}}

   \institute{$^1$ Instituto de Astrof\'\i sica de Canarias, 
E-38205 La Laguna, Tenerife, Spain\\
$^2$ Departamento de Astrof\'\i sica, Universidad de La Laguna,
E-38206 La Laguna, Tenerife, Spain}

   \date{Received xxxx; accepted xxxx}

 
  \abstract
  {}
   {Dust in spiral galaxies produces emission in the far-infrared (FIR) and internal absorption in visible wavelengths. However, the relation of the two amounts is not trivial because optical absorption may saturate, but the FIR emission
does not. Moreover, the volume concentration of dust plays a role in the relation of absorption and emission, which depends on the size of the galaxy. We explore
the relation of these three quantities.}
   {In order to understand the geometrical problem, we developed a model of dust distribution.
We also investigated the relation of the three variables with real data 
of spiral galaxies at $z<0.2$ using the spectroscopic Sloan Digital Sky Survey (SDSS) and FIR AKARI survey. Internal absorptions were derived with two different methods: the ratio of emission lines
H$_\alpha $ and H$_\beta $, and a previously calibrated relation based on the color variations
as a function of absolute magnitude and concentration index.}
   {We find that in our low-$z$ sample, the dependence of the average internal attenuation
on galaxy size is negligible on average because of the relation of dust mass with size. It allows us to derive the internal attenuation of the galaxy, $A_V$, even when we only know its FIR flux. 
This attenuation approximately depends on the inclination of the galaxy $i$ as
$\overline {A_V}=\overline {\gamma_V} \log _{10}\left(\frac{1}{\cos i}\right)$, where 
$\gamma_V$ is a constant. 
We found that $\gamma_V$ has a maximum value for galaxies 
 of $1.45\pm 0.27$ magnitudes.\\
When similar properties of dust are assumed, a general expression can be used at any $z$:
$\overline {\gamma_V }=(1.45\pm 0.27)f_M^{\exp{[-(1.0\pm 0.6)f_M]}}$ and
$f_M=7.6\times 10^{-6} \alpha_{hR}^{-1.75}\times
\left(\frac{F_{\rm FIR}}{700\ {\rm Jy}}\right)^{1.87}
\times f_{\rm cosmol.}(z)$; the dependence on the cosmological model is embedded in
$f_{\rm cosmol.}(z)=d_L(z)({\rm Mpc})^2(1+z)^{(1.75\eta -1.87)}$, where
$\eta =2$ for cosmologies following Etherington's relation, $d_L$ is the luminosity distance, 
$\alpha _{hR}$ is the angular size
of the scalelength, and $F_{\rm FIR}$ the flux at wavelength $100(1+z)\ \mu $m.}
   {For cases of nonsaturation ($f\lesssim 3.6$), this might be used as a cosmological test because the factor $f_{{\rm cosmol.}}$ at high $z$ 
varies strongly in different cosmologies.
Although the present-day sensitivity of FIR or millimeter surveys does not allow us to carry out this
cosmological test within the standard model, it may be used in the future, when we can observe  galaxies at $z=3-5$ with a sensitivity at $\sim $500 $\mu $m better than $\sim 10\ \mu$Jy, for instance. For much lower $z$ or 
different cosmological models, a test might be feasible at present.}


\titlerunning{Absorption, dust emission, and galaxy size}
\authorrunning{L\'opez-Corredoira \& Guti\'errez}

   \maketitle
%

\section{Introduction}
\label{.intro}

The distribution of dust and its corresponding absorption are well known in the Milky Way 
\citep[e.g.,][]{Una98,Dri03,Sch11} and in other nearby spiral galaxies \citep[e.g.,][]{Pel95,Pal97,Cho09,Poh10,Dev16,Vie17,Bia18,Sal18,Sal20}. 
Attenuation curves result from a combination of dust grain properties, dust content, and the spatial arrangement of dust and different populations of stars \citep[e.g.,][]{Sal20}.
Maps of Galactic extinction and internal attenuation within a galaxy are important 
to correct the magnitudes of extragalactic objects, in order to derive the properties 
of the stellar distribution in that galaxy.

Although the terms extinction and attenuation are sometimes used interchangeably in
the literature, the internal absorption we analyzed refers more to the usual meaning of the second term.
Dust attenuation refers to the general effect on the spectrum of an extended object 
through dust; in this sense,
all extinction curves are attenuation curves, but not all attenuation curves are extinction curves.
In general, 'attenuation' is used here to indicate that the geometry of the sources and dust in a system is more complex than a single point-like object in the background of a dust layer. The measurement and/or 
theoretical calculation of attenuation is therefore significantly more complex than for extinction because
it also includes the effects arising from the distribution of stars and dust
in the galaxy \citep{Sal20}.

There are different methods to derive the internal attenuation in spiral galaxies that are different from the Milky Way.
The simplest method is using the colors and concentration indexes \citep[e.g.,][]{Cho09}, or, when many filters
are available, the fitting of photometry in several bands, the spectral energy distribution (SED), 
using templates based on stellar population synthesis models \citep[e.g.,][]{Bru03,Mar05,Vaz10} to model the stellar emission, which is perhaps the most common 
method at either low z \citep[e.g.,][]{Boq13,Dec19} or high $z$ \citep[e.g.,][]{Wal11,Con13,LoF17,Bua18,Sch18}. The attenuation curve can also be derived by modeling the galaxy SEDs using radiative transfer 
calculations \citep[e.g.,][]{DeL14,Ner20}.
Another way, applicable to galaxies with high ratios of star formation or to active galaxies, would
be using the ratio of some spectral lines with known intrinsic values, for instance, the ratio of $H_\alpha $
and $H_\beta $ \citep{Cal01}. 

Dust also produces emission, and this is another way to investigate its distribution, by using
far-infrared (FIR) surveys \citep[e.g.,][]{Cal13,Lam19}. Because the dust that produces FIR emission
and internal attenuation is the same, it is clear that the  two quantities must be related.
In principle, we would expect a linear proportionality, but it is not so simple because optical absorption may saturate, that is, over a given amount of dust, we cannot see the stars in the far part of the galaxy, and adding more dust will not add more extinction over these stars, but the FIR emission
does not saturate because the FIR absorption is totally negligible and all the radiation of the dust in a
galaxy comes to the observer, regardless of how much dust we put on it.

Another element that plays a role in the relation of
absorption and emission is the volume concentration of dust,
which depends on the size of the galaxy.  A galaxy with a given amount of dust
and a large diameter receives a low average density of dust that produces small absorption. The opposite
is also true: When
the same amount of dust that produces the same FIR luminosity is distributed within a galaxy with a very small diameter, the concentration of dust is much higher, and consequently, the average absorption is much larger.

Qualitatively, the relation among the three variables (internal attenuation, dust emission, 
and size of spiral galaxies) is clear, but the quantitative evaluation of this relation is not immediate and needs some analyses. This is precisely one of the purposes of this paper: modeling the dust distribution to understand the
connection among the three variables, which will be later calibrated with real data from the local universe.
For modeling the dust distribution that gives rise to the galactic attenuation, the
literature is rich: either based on analytic prescriptions for the galactic 
structure \citep[e.g.,][]{Wit00,Ino06,Pop11,Seo16}, or semianalytical prescriptions \citep{Gra00,Fon11,Wil12,Gon13,Pop17}, or 
complex hydrodynamic galaxy formation simulations \citep{Jon06,Roc08,Nat15,Nar18,Tra20}, many of them
including a cosmological context. We do not enter in the discussion of complex models, but use
simple analytical expressions, and we mainly focus on the statistical relation among the three variables, with the purpose of obtaining some 
easy-to-use recipes that can be used with multiple purposes, such as obtaining the average expected value of one of the variables when we know the other two.

The dependence of the absorption on the galaxy size and the dust mass (obtained from FIR emission) 
is indeed interesting as a cosmological test of the galaxies at high $z$.
For instance, it has been questioned whether the strong linear size evolution of spiral galaxies \citep{Tru06,Shi15} is intrinsic or is an artifact due to the use of an incorrect cosmological model \citep{Lop10,Ler18,Bal19}.
A tool for these characteristics might therefore break the degeneracy and give an answer, provided that
the dust characteristics do not change the temperature and emission and absorption properties with time.

In order to understand the geometrical problem, we develop a simple toy model in \S \ref{.toy}.
The real relation for low-redshift ($z<0.2$) galaxies is obtained for a sample derived
from a cross-correlation of the optical Sloan Digital Sky Survey (SDSS) spectroscopic survey \citep{Yor00,Aba09} 
and the FIR AKARI survey \citep{Doi15} in \S \ref{.local}.
In \S \ref{.highz} we indicate how this relation among attenuation, FIR emission, and size can be extrapolated at high $z$ as a possible cosmological test.

\section{Toy model of internal dust absorption in spiral galaxies}
\label{.toy}

\subsection{Basic considerations}

First, we give the basic equations that relate the average attenuation in magnitudes with the distribution
of flux in the galaxy.
Within a maximum angular radius $\alpha $ from the center of the galaxy, corresponding to a linear size $R_{0,max}=\alpha d_A$, where $d_A$ is the angular distance of the galaxy, the average internal attenuation in V-band measured by an external observer is
\begin{equation}
\overline{A_V}=\frac{\int _0^{2\pi} d\theta _0\int _0^{R_{0,max}}dR_0\,R_0A_V(R_0,\theta _0)F_{V,obs}(R_0,\theta _0) }
{F_{V,obs,total}}
,\end{equation}
where $R_0$ and $\theta _0$ are polar coordinates of the linear projected area onto the sky; $F_{V,obs}(R_0,\theta _0)$, $A_V(R_0,\theta _0)$ are the corresponding observed flux after extinction absorption and cumulative extinction in those coordinates, respectively; and $F_{V,obs,total}$ is the total integrated flux in the area. The attenuation is related to the flux through
\begin{equation}
\label{avr0t0}
A_V(R_0,\theta _0)=2.5\log_{10}\frac{F_{V,em}(R_0,\theta _0)}{F_{V,obs}(R_0,\theta _0)}
,\end{equation}
where $F_{V,em}$ is the emitted flux of the stars in V band (before attenuation correction). 


Because the galaxies have different inclinations $i$ and their internal absorption is proportional to
$\log _{10}\left(\frac{1}{\cos i}\right)$ \citep{Sha07,Cho09} [where in practice $\cos i\approx \frac{r_{\rm minor}}{r_{\rm major}}$, with $r_{\rm major}$ and $r_{\rm minor}$ the major and minor axis of the projected disk galaxy] , we would instead consider the amount $\gamma _V$ independent of the inclination, 
which is defined as  
\begin{equation}
\label{gammaV}
\gamma _V=\frac{\overline {A_V}}{\log _{10}\left(\frac{1}{\cos i}\right)}
.\end{equation}

\subsection{Disk model}

Previous equations can be used to compute the attenuation for any flux and dust distribution of any galaxy.
We applied them to the particular case of a spiral galaxy with an exponential disk in the stellar populations
and dust. It does not include spiral arms, a stellar halo, or the central components (bulge, long bar, or stellar ring).  Therefore this simple toy model may better represent disk-dominated galaxies, in which the ratio of bulge flux to the total is low. Nonetheless, for the relation of Petrosian radii and scalelength in \S \ref{.petro}, we show that the effect of the bulge is negligible, and for the distribution of stars and dust, the effect of
the bulge changes the distribution of stars and dust in the central parts only slightly, as we show in 
\S \ref{.bulge}.  The relations derived in this section
should be interpreted as a rough approximation of the expected behavior, not as an accurate predictive model, which is better derived in the calibration with real galaxies in \S \ref{.local}.

For an axisymmetric star and dust distribution corresponding to a disk,
\begin{equation}
F_{V,em}(R_0,\theta _0)=\int _{-\infty}^\infty dr \rho _{*,V}(R,z)
,\end{equation}\[
R=R_0\sqrt{\cos ^2\theta_0 +\frac{\sin ^2\theta _0}{\cos ^2i}},\]
\[z=R\,\cos i,\]
\begin{equation}
F_{V,obs}=\int _{-\infty}^\infty dr\, \rho _{*,V}(R,z)\,
\exp{\left[-\kappa _V\int _r^\infty dr'\rho_d(R,z=r'\cos i )\right]}
,\end{equation}      
where $R$, $z$ are the coordinates in the plane of the galaxy, $r$ is the distance from the plane of the galaxy along the line of sight with positive values toward the observer, $\rho _{*,V}$ is the emitted flux per unit volume in V band by the stellar component, and $\rho _d$ is the dust density.

For a simple exponential disk for stars and dust, with the same scalelength $h_R$ for both components,
\begin{equation}
\label{rhodisc}
\rho _{*,V}=\frac{F_{V,total,em}}{4\pi h_R^2h_{z,*}}\exp{\left[-\frac{R}{h_R}\right]}\exp{\left[-\frac{|z|}{h_{z,*}}\right]}
,\end{equation}
\begin{equation}
\label{rhodustdisc}
\rho _d=\frac{M_d}{4\pi h_R^2h_{z,d}}\exp{\left[-\frac{R}{h_R}\right]}\exp{\left[-\frac{|z|}{h_{z,d}}\right]}
,\end{equation}
where $M_d$ is the total mass of dust in the exponential disk.
Hence, with Eq. (\ref{avr0t0}),
\begin{equation}
A_V(R_0,\theta _0)=2.5\,
\log_{10}\frac{(2h_{z,*}/\cos i)}{H_0+H_1}
,\end{equation}\[
H_j=\int _0^\infty dr \exp{\left[-\frac{r\,\cos i}{h_{z,*}} \right]} \]\[
\ \ \ \ \times \exp{\left[ -\frac{ \kappa _VM_d   \exp{\left[-\frac{R}{h_R}\right]}\left(2*j+(-1)^j\exp{\left[-\frac{r\cos i}{h_{z,d}} \right]}\right)}{4\pi h_R^2\cos i} \right]}
.\]
With this model, $\gamma _V$ is independent of the absolute magnitude of the galaxy because 
the amplitude $F_{V,total,em}$ cancels out in numerator and denominator of Eq. (\ref{avr0t0}).

This simple model fails in representing the flux of spiral galaxies in the central part, where
a bulge or bar may produce a high amount of light, and where the disk usually presents a deficit of
stars with respect to a pure exponential disk \citep{Lop04}. Nonetheless, the deficit of flux produced by a
hole in the disk may roughly be compensated for with the excess produced by the other central components; and
for the external part, the exponential disk is a quite accurate representation.
 
For low-attenuation cases ($A_V<<1$), $A_V\approx 1.086\,\kappa _V\,M_d\,g(R,h_R,h_{z,*},h_{z,d})$, that is, proportional to the total dust mass and a function $g$ depending on the geometrical scales of the disk.
For low attenuation, the ratio $\frac{\gamma _V}{M_d}$ is therefore independent of the dust mass  and only depends on the geometrical distribution of stars and dust. However, as we show below, for
high dust masses, the relation is not linear.

\subsection{Numerical example}

We carried out a numerical calculation for some usual values of the parameters.
We set $\kappa _V=1.33\times 10^3$ m$^2$ kg$^{-1}$ \citep{Loe97}.
We set the scalelength $h_R$, and dust mass $M_d$ variable parameters around typical values
of the Milky Way: $h_R=3$ kpc \citep{McM11} and $M_d=M_{d,MW}=2.8\times 10^7$ M$_\odot $ \citep{Gut14}. 
We assumed the average value of $\cos i=0.5$.

\begin{figure*}[htb]
\vspace{0cm}
\centering
\includegraphics[width=5.8cm]{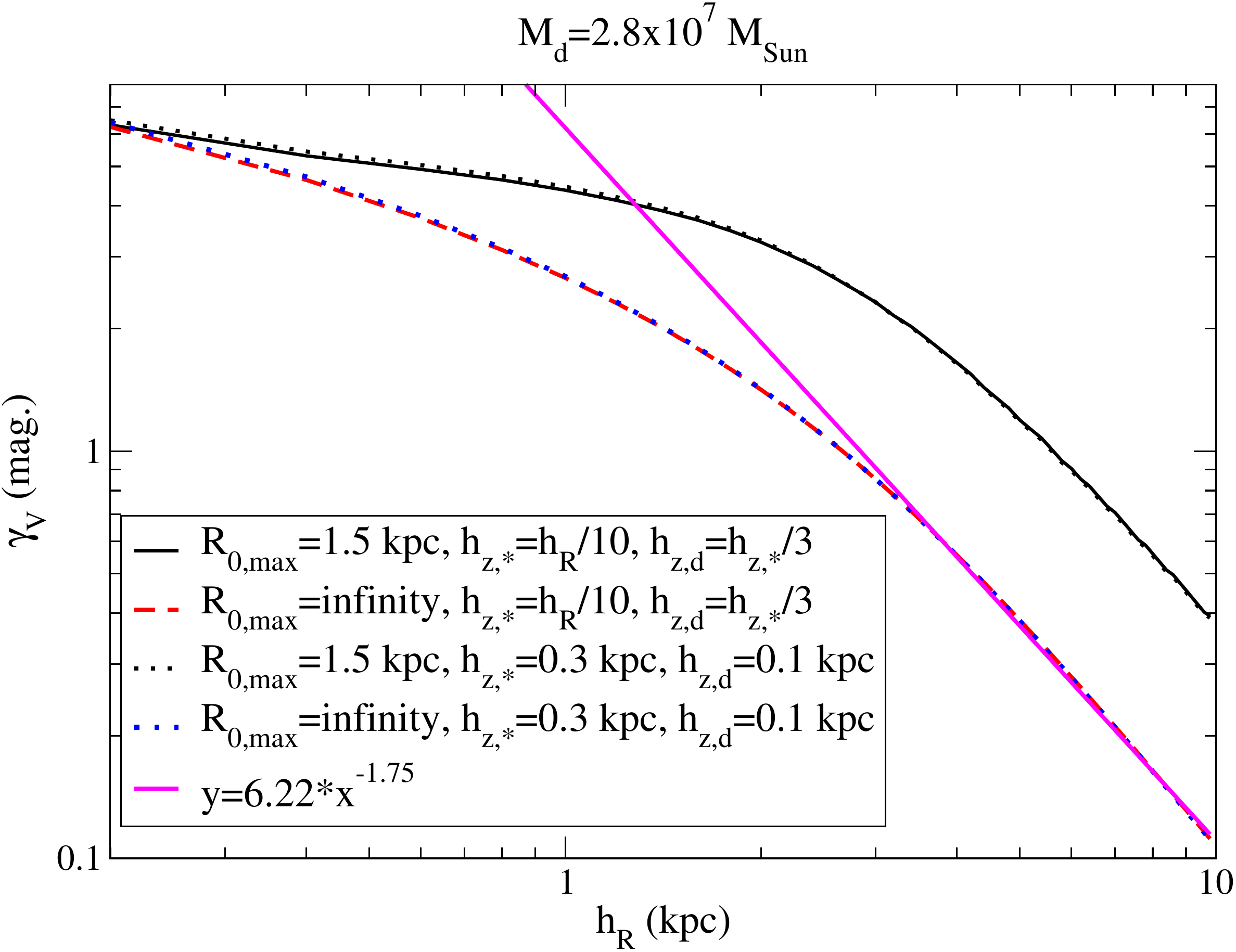}
\hspace{.2cm}
\includegraphics[width=5.8cm]{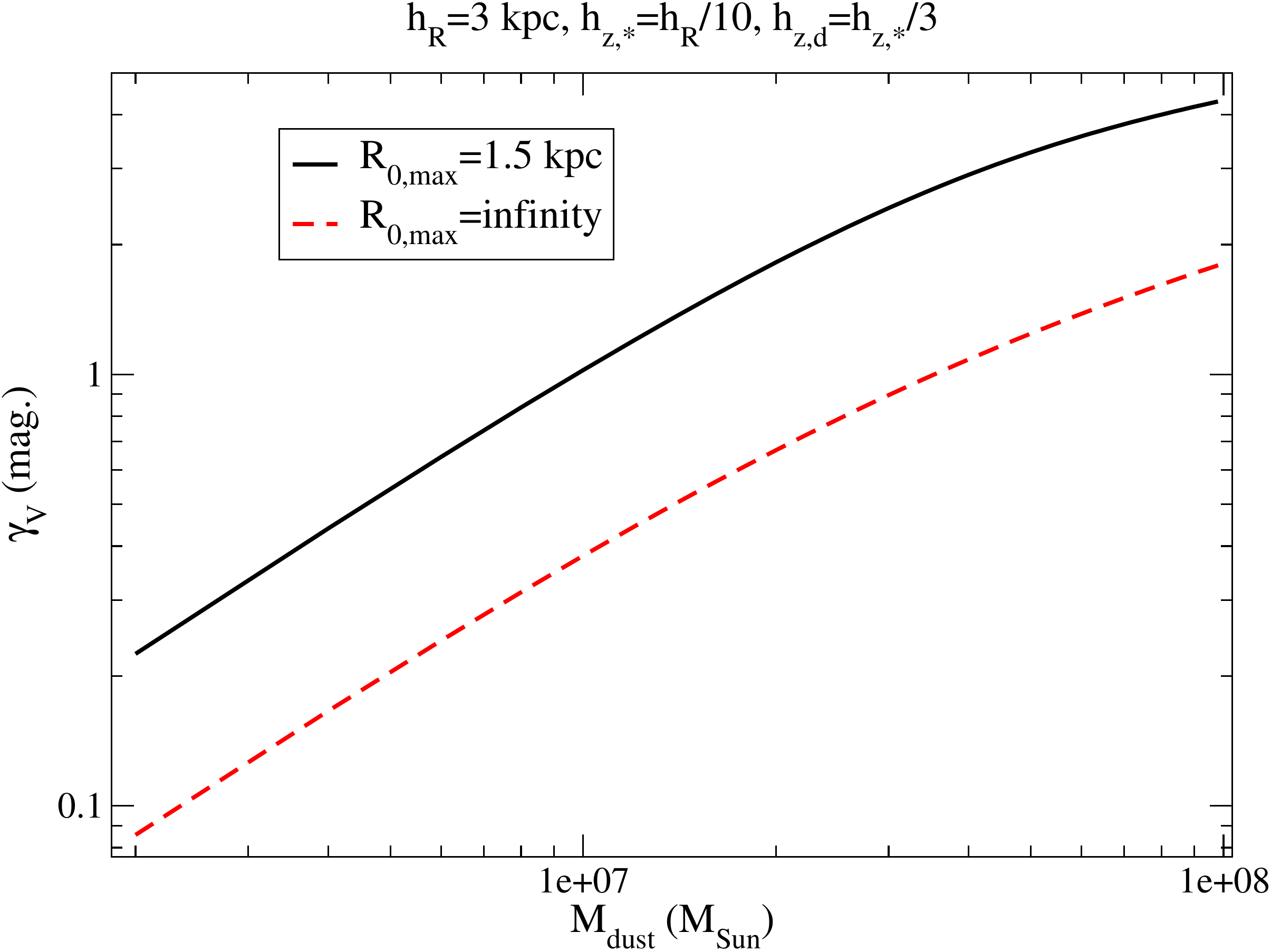}
\hspace{.2cm}
\includegraphics[width=5.8cm]{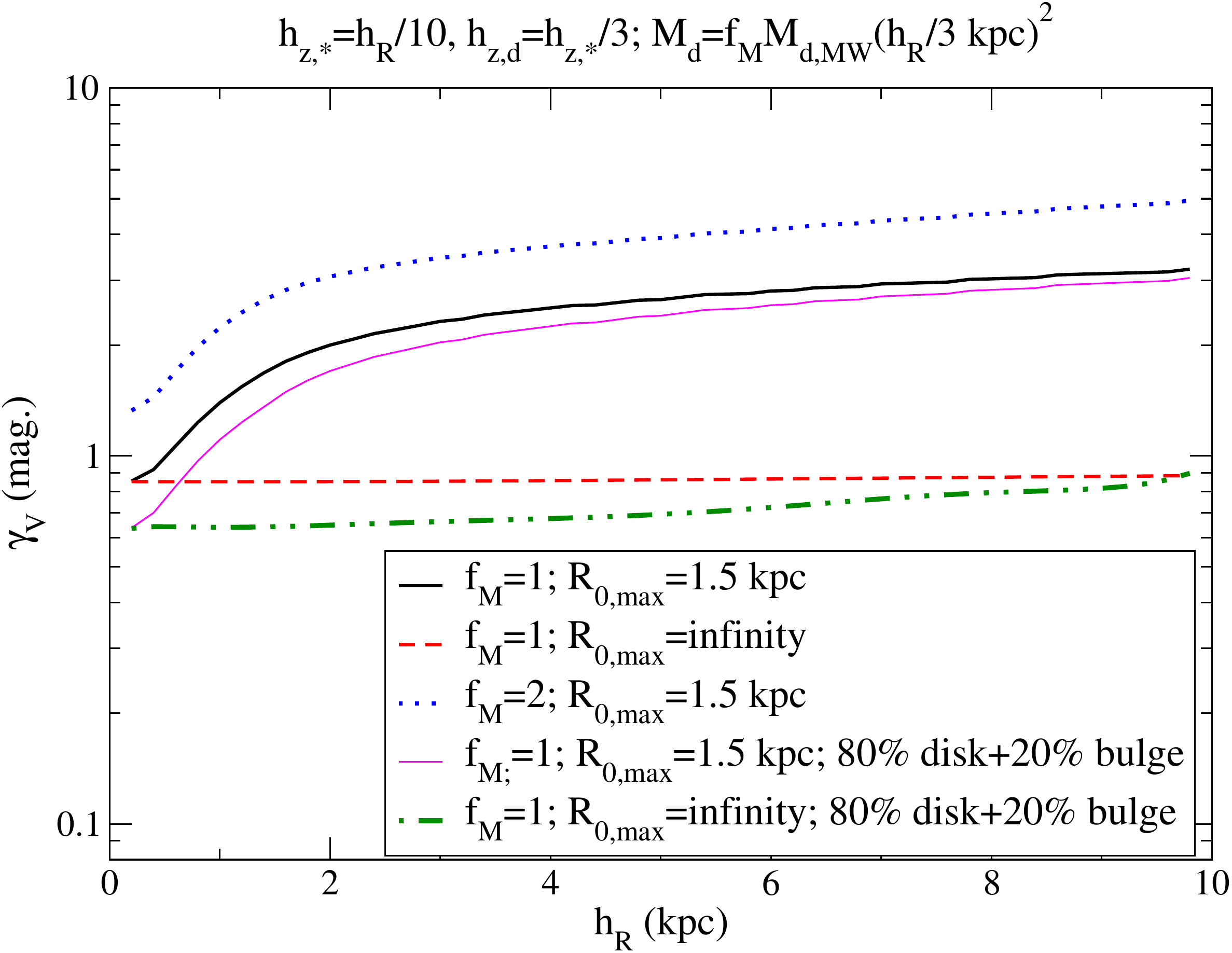}
\caption{Theoretical value of $\gamma _V$ for a model of a purely exponential disk in three cases. Left: constant
dust mass. Middle: Constant scaleheight. Right: Dust mass proportional to the square of the scalelength.}
\label{Fig:gamma_hr}
\end{figure*}

First, we assumed an angular radius $\alpha =1.5"$ corresponding to the radius of fibers for spectroscopy
in the SDSS survey \citep{Yor00} and a typical angular distance of 200 kpc (corresponding to $z=0.050$ with the
standard cosmology, $h_0=0.7$, $\Omega _\Lambda=0.7$), that is, $R_{0,max}=1.5$ kpc.
 We plot the average absorption $\gamma _V$ (absorption normalized to $\cos i=0.1$)  versus the
scalelength $h_R$ for different combinations of scaleheights (typical values obtained in the Milky Way) in the left panel of Fig. \ref{Fig:gamma_hr}.
Second, we assumed $R_{0,max}=\infty$, that is, the average attenuation of the whole galaxy. We plot in the left panel of
Fig. \ref{Fig:gamma_hr} $\gamma _V$ versus scalelength, also for different combinations of
scaleheights. For any value of $R_{0,max}$, the mean
absorption decreases with the size of the galaxy (proportional to $h_R$), but it is almost independent
of the variation in scaleheight. For low $h_R$, the absorption saturates and goes asymptotically to a limit.
With lowest $R_{0,max}$, the mean absorption is higher because we  averaged only in the central parts of the galaxy.
In the central panel of Fig. \ref{Fig:gamma_hr} we plot this for a fixed $h_R$ and as a function of the dust mass, where the relation is linear for low masses, but slowly grows for high masses.
The ratio of the average absorption up to $R_{0,max}=1.5$ kpc and the average
total absorption remain more or less constant. Therefore we can apply a 
correcting factor $C(1.5\ {\rm kpc})\equiv 
\frac{\gamma _V(R_{0,max}=\infty)}{\gamma _V(R_{0,max}=1.5\ {\rm kpc})}$ in the absorption that only depends on the size of the galaxy
to take the limited size of the SDSS fiber into account,
\begin{equation}
\label{factorC}
\log _{10} C(1.5\ {\rm kpc})\approx -0.24-0.34\log_{10} h_R\ {\rm (kpc)}
,\end{equation}
derived from a fit in the left panel of Fig. \ref{Fig:gamma_hr} (r.m.s. of the fit with respect to $C$ equal to $0.052$).

Although the dependence of the absorption and dust mass in the left and central panel of Fig. \ref{Fig:gamma_hr}
is clear, these two numerical calculations are not realistic in practice because $h_R$ and $M_d$ are not
independent variables. 
A more realistic case would be modeling the dependence of $M_d$ on $h_R$ as well. We assumed for
a third experiment $M_d\propto h_R^2$ (this dependence is indeed observed in the data; see \S \ref{.local}), that is, we made $\rho _d(R=0,z=0)$ constant (see Eq. (\ref{rhodustdisc})).
The dependence of $\gamma _V$ on $h_R$ (proportional to the square root of $M_d$) is illustrated
in the right panel pf Fig. \ref{Fig:gamma_hr}.
The net increase of the average absorption with scalelength for $R_{0,max}=1.5$ kpc and
the constant average absorption when integrated in the whole galaxy ($R_{0,max}=\infty $) are interesting results.

\subsection{Adding a bulge}
\label{.bulge}

The previous toy model only includes an exponential disk of stars and dust. 
With the same method, a set of models might be created in which we can substitute Eqs. (\ref{rhodisc}) and
(\ref{rhodustdisc}) for other more complex expressions, including a clumpy disk and/or a bulge. The results will change, but the trends of the variation in attenuation with size and dust mass will be similar.
For instance, we substituted Eq. (\ref{rhodisc}) for 
\begin{equation}
\rho _{*,V}=\left \{ 
\begin{array}{ll}
K_1(h_R)\exp{\left[-\sqrt{R^2+(2.5z)^2}/(2.5hr)\right]},& \mbox{$R<h_R$} \\
K_2(h_R)\exp{\left[-\frac{R}{h_R}\right]}\exp{\left[-\frac{|z|}{h_{z,*}}\right]},& \mbox{$R\ge h_R$} 
\end{array}
\right \} \;,
\end{equation}
that is, an exponential disk for $R\ge h_R$ and a bulge (taking the expression by \citet{Lop05} for the
Milky Way ellipsoidal bulge, but not considering the triaxiality and setting an axisymmetry), and setting
$K_1(h_R)$ and $K_2(h_R)$ such that the disk contributes 80\% of the observed light, and the bulge
contributes 20\%. The result for the case of $M_d\propto h_R^2$ is shown in the right panel of Fig. \ref{Fig:gamma_hr}. 
We observe variations with respect to the disk alone, but with similar trends.

\subsection{Relation of scalelength and Petrosian radii}
\label{.petro}

In the SDSS, Petrosian radii are used to measure the size of the galaxies instead of the scalelength, but
both quantities can be easily related.
The SDSS provides the Petrosian
radius, defined as $RP$ such that \citep{Shi01}\footnote{http://spiff.rit.edu/classes/phys443/lectures/gal\_1/petro/petro.html}
\begin{equation}
\frac{1}{2\pi }\int _0^{2\pi }d\theta _0 F_{V,obs}(RP,\theta _0)=
0.2\overline {F_{V,obs}}(RP)
,\end{equation}\[
\overline {F_{V,obs}}(RP)=\frac{1}{\pi \,(RP)^2}\int _0^{2\pi} d\theta _0\int _0^{RP}dR_0\,R_0F_{V,obs}(R_0,\theta _0) 
.\]
The Petrosian flux used in the SDSS is
\begin{equation}
FP\equiv \overline {F_{V,obs}}(2\,RP)
.\end{equation}
The Petrosian half-light and 90\% light radii are $RP_{50}$ and $RP_{90}$ , respectively, such that
\begin{equation}
\overline {F_{V,obs}}(RP_{50})=0.5\,FP
,\end{equation}
\begin{equation}
\overline {F_{V,obs}}(RP_{90})=0.9\,FP
.\end{equation}

\begin{figure}[htb]
\vspace{0cm}
\centering
\includegraphics[width=9cm]{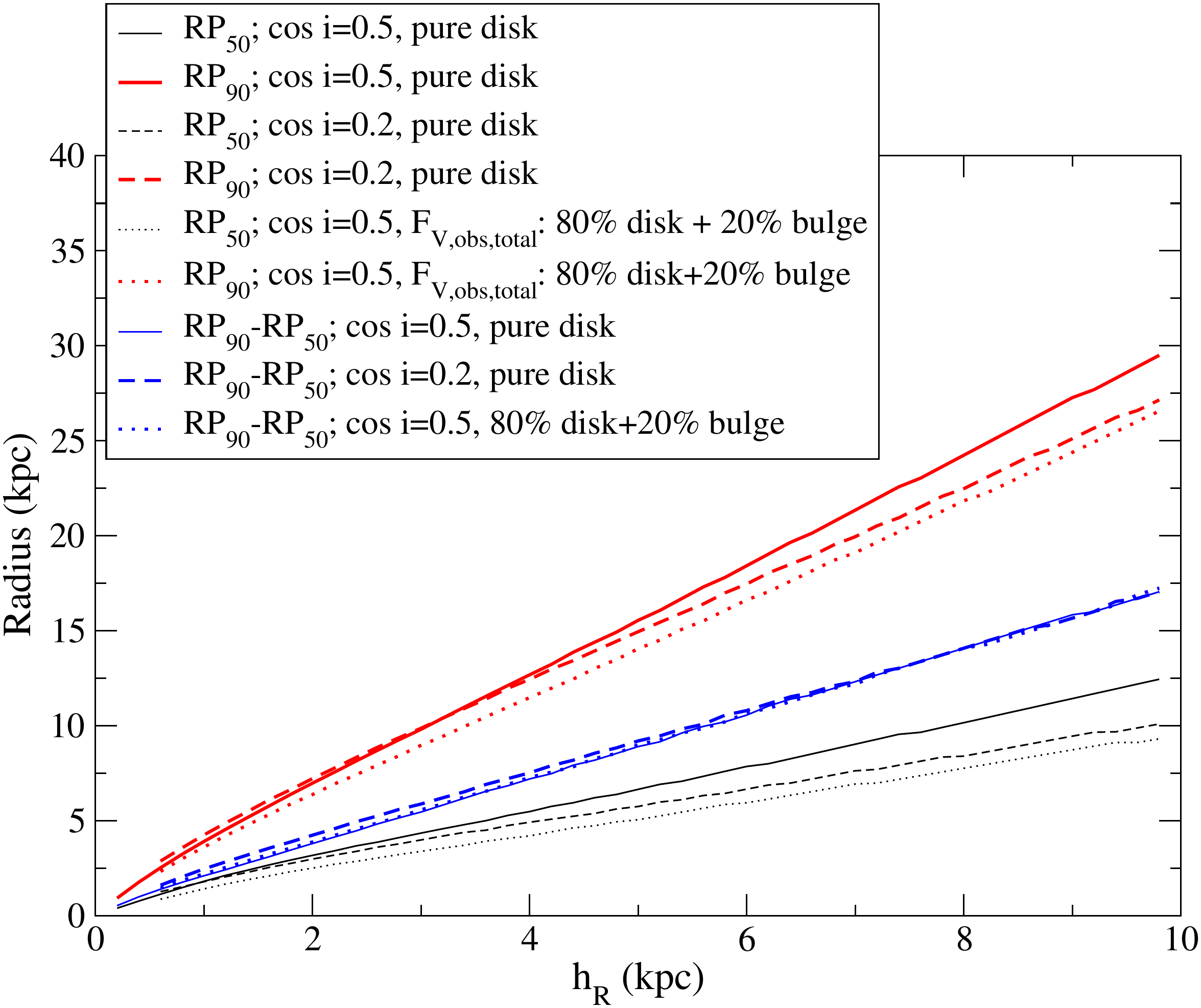}
\caption{Relation of Petrosian half-light and 90\% light radii with the scalelength of a spiral galaxy, either with a purely exponential disk or with some bulge component with 20\% of the light; we set $h_{z,*}=0.3$ kpc, $h_{z,d}=0.1$ kpc, and $M_d=2.8\times 10^7$ M$_\odot $.}
\label{Fig:petros}
\end{figure}

When we apply these definitions to our previous disk model, we obtain the results that we show in Fig. \ref{Fig:petros}.
In the same figure, we also show $RP_{50}$ and $RP_{90}$ of the cases with a different inclination of the galaxy ($\cos i=0.2$ instead of 0.5) and including a bulge component that is represented by a ratio of the total observed light coming from the center of the galaxy apart from the disk. The three different cases for $RP_{50}$ and $RP_{90}$ show slight differences. However, when we calculate $RP_{90}-RP_{50}$, this
last amount as a function of $h_R$ is almost independent of the inclination of the galaxy and the ratio of the bulge contamination. This is expected because the outer light between 50\% and 90\% only traces the disk.
Therefore we took this amount to determine the correlation with $h_R$ independently of other parameters. A linear fit of $RP_{90}-RP_{50}$ in Fig. \ref{Fig:petros} for $\cos i=0.5$ and a pure disk gives
\begin{equation}
\label{scale_petro}
RP_{90}-RP_{50}=0.35\ {\rm kpc}+1.71\,h_R
.\end{equation}
The deviation from a perfect proportionality ($[RP_{90}-RP_{50}]\propto h_R$) stems from the nonzero thickness of the disk. The term 0.35 kpc is almost negligible and does not affect our calculations strongly. We use this relation to derive $h_R$ from $RP_{50}$ and $RP_{90}$ given in the SDSS data, independently of the inclination and the bulge ratio.

\section{Local calibration of absorption as a function of size and dust emission}
\label{.local}

The theoretical calculations in the previous section give us a clear relation of absorption, dust emission (correlated with the dust mass), and size of the galaxy. We assumed a simplistic
model, however, and the reality may be more complex. Particularly, the dependence on the size of the galaxy depends on the geometrical distribution of gas and stars. In order to determine this relation
in reality, we calibrated it with local ($z<0.2$) galaxies using a cross-correlation of the surveys SDSS and AKARI for FIR fluxes. 

\subsection{Data}

For our purpose, we need information of the galaxy sizes and colors and/or spectra, which we obtained from the SDSS. We also need FIR fluxes to calculate dust masses, which we obtained using the AKARI survey.

Photometric and morphological information in the optical range was obtained from the SDSS Data Release 7 (SDSS-DR7; \citet{Yor00,Aba09}). 
The spectroscopic data were taken from SDSS-Data Release 14 \citep{Abo18}.\footnote{File specObj-dr14.fits available at https://www.sdss.org/dr14/spectro/spectro\_access/}
Petrosian radii $RP_{50}$ and $RP_{90}$, magnitudes, fluxes in $H_\alpha $ and $H_\beta $ 
, and the galaxy classification were taken from the SDSS data as well.
The scalelength $h_R$ was derived with Eq. (\ref{scale_petro}).
Data for the inclination $i$ were obtained from the ellipticity $\epsilon $ of the projected galaxy, in which
$\cos i=1-\epsilon $, and $\epsilon=\frac{1-\sqrt{U^2+Q^2}}{1+\sqrt{U^2+Q^2}}$, with $U$, $Q$ the Stokes parameters given by the SDSS (we took them at $r$ band).
We selected only galaxies with significant emission lines of H$_\alpha $ and H$_\beta $: $>3\sigma $ detection.

AKARI \citep{Mur07,Doi15} is a FIR all-sky atlas from a sensitive all-sky survey using the Japanese AKARI satellite. The survey covers $>99$\% of the sky in four photometric bands centered at 65 $\mu $m, 
90 $\mu$m, 140 $\mu $m, and 160 $\mu $m, with spatial resolutions ranging from 1.0 to 1.5 
arcseconds. It has
a better spatial resolution and wider wavelength coverage than previous all-sky survey such as
IRAS. AKARI was operated with a telescope with a 68.5 cm diameter, cooled down to 6 K and observed from a Sun-synchronous polar orbit at 700 km altitude. It was successfully launched on 21 February 2006 by an M-V rocket from the Uchinoura Space Center, Japan. AKARI ran out of its onboard supply of cryogen, liquid helium, on August 26, 2007, after successful operation and observations that began on May 8, 2006, achieving the expected lifetime of 550 days, including the performance verification phase and three observation phases. This survey was particularly useful for exploring the dusty Universe.
We only used data of galaxies with flags GRADE and FQUAL90 of AKARI with a maximum quality equal to 3 and available fluxes in the four wavelengths.

\begin{figure*}[htb]
\vspace{0cm}
\centering
\includegraphics[width=16cm]{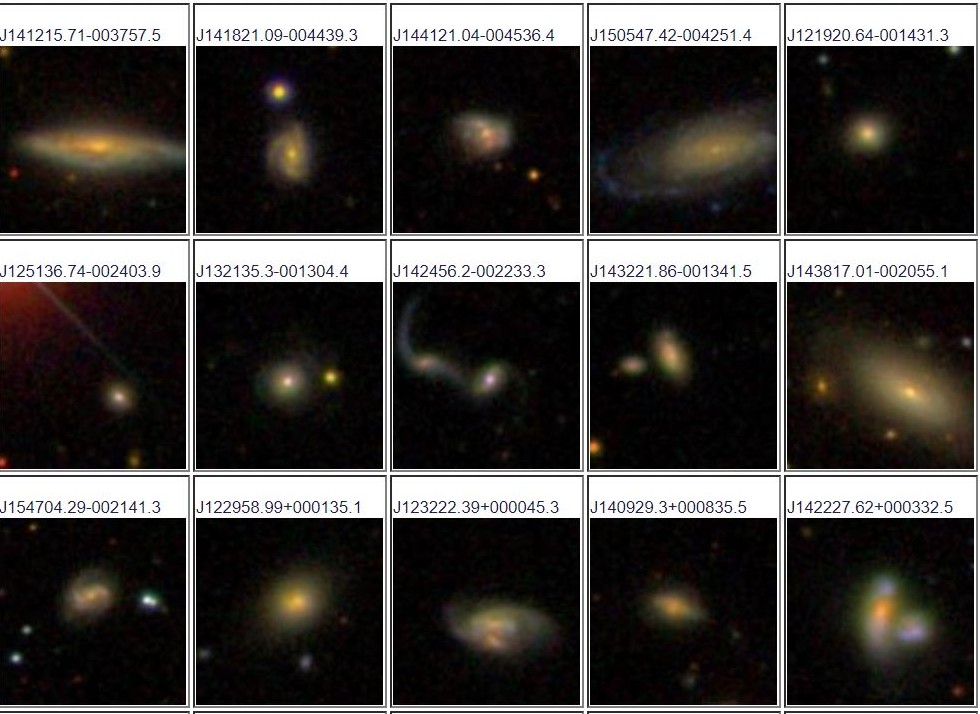}
\caption{Examples of SDSS-selected galaxies with a counterpart in AKARI, that is, with a significant amount of
dust.}
\label{Fig:examples}
\end{figure*}

First, we cross-correlated the AKARI data with those from the SDSS, matching objects of the AKARI and SDSS catalogs with separations $<10$ arcseconds. This resulted in a sample of 9,059 galaxies. Then we completed the required information by selecting only galaxies with measured H$_\alpha $ and H$_\beta $ flux in the SDSS galspecline catalog. Finally, for morphology, we used the classification made by \citet{Mee15}, which provides 2D decompositions in the Sloan g, r, and i bands for several parametric models (de Vaucouleurs, S\'ersic, de Vaucouleurs plus exponential disk, and S\'ersic plus exponential disk). The final sample containing photometric, spectroscopic, and morphology information comprises 5,386 galaxies.
Examples of the selected galaxies are shown in Fig. \ref{Fig:examples}.
The use of nearby galaxies with $z<0.2$ makes the angular and luminosity distance calculation almost independent of the cosmological model, that is, with a dependence on the cosmological model that is negligible. We calculated these distances from the redshift using the standard $\Lambda $CDM model with $h_0=0.7$, $\Omega _\Lambda =0.7$.

The selected galaxies have significant amounts of dust so that they can be detected in the FIR regime by AKARI. In addition, all galaxies have H$_\alpha $ and H$_\beta $ emission lines, which means that they should also be star forming. Therefore these are spiral
or elliptical galaxies with a high amount of dust that might be produced by residual star formation \citep{Lop17}.
About 1-2\% of the whole set of galaxies were morphologically classified by the SDSS as ellipticals.
However, all of them present significant emission lines, which means that they were  either misclassified based on their apparent morphology, or they are elliptical galaxies with a high amount of residual star formation and/or nuclear
activity.

%
%
%
%

\subsection{Relation of dust mass to FIR at rest luminosities}

In our model in \S \ref{.toy} we used the dust mass instead of the FIR luminosities given by AKARI,
but they can be approximately related to each other.
We fit the observed fluxes $F_{\nu, {\rm obs}}$ in the four AKARI wavelengths $\lambda $ to a dependence \citep[e.g.,][\S 4]{Lop17b}
\begin{equation}
F_{\rm em}(\nu _{\rm rest})=\frac{F_{\nu ,{\rm obs}}}{(1+z)}=A_d\,\nu_{\rm rest}^{\beta _d}\,B(\nu _{\rm rest},T_d)
,\end{equation}
\[\nu_{\rm rest}=\frac{c\,(1+z)}{\lambda },\]
where $T_d$ is the dust temperature, $\beta _d$ is the emissivity, $A_d$ is the amplitude, and
$B(\nu ,T)$ is the blackbody emission function. We obtain the parameters $T_d$, $\beta _d$, $A_d$ with the
four available frequencies, taking the error bars of the flux into account, in a $\chi ^2$ fit. 
After we derived the three parameters of the dust emission, we obtained the bolometric luminosity through

\begin{equation}
L_{\rm bol,dust}=4\pi\,A_dd_L(z)^2
\int _0^\infty d\nu\,\nu ^{\beta _d}B(\nu ,T_d) 
,\end{equation}
where $d_L(z)$ is the luminosity distance. In the galaxies with an error in $L_{\rm bol,dust}$ lower than 20\%, 
the average
and r.m.s. of the two free parameters are $\langle \beta \rangle =2.4$, $\sigma _\beta =2.1$,
$\langle T_d \rangle =23$ K, and $\sigma _{T_d} =10$ K.

The calibration with mass can be made with the Milky Way dust, for which 
$L_{\rm bol,MW}=2.6\times 10^{43}$ erg/s \citep{Dav97}, and the dust mass is $M_{d,MW}=2.8\times 10^7$ M$_\odot $.
Therefore
\begin{equation}
\label{md}
M_d\approx \frac{L_{\rm bol,dust}}{2.6\times 10^{43}\ {\rm erg/s}} M_{d,MW} 
.\end{equation}

\subsection{Method 1 for calculating the internal absorption}

A first method for calculating the absorption is using the ratio of the fluxes of the $H_\alpha $ and
$H_\beta $ emission lines in the spectra of the galaxies: $F_{H_\alpha }$ and $F_{H_\beta }$, respectively. 
The spectra are available in the SDSS, although with
a fiber diameter of 3", which assumes that the area in which we average
the absorption is restricted. For an average $z=0.05$ galaxy, we would be observing the innermost 1.5 kpc of the galaxy.
The average attenuation is \citep{Cal01}

\begin{equation}
\gamma _V=\frac{\overline {A_V}}{\log _{10}\left(\frac{1}{\cos i}\right)}
,\end{equation}\[
\overline{A_V}=11.6\left[\log_{10}\left(\frac{F_{H_\alpha}}{F_{H_\beta }}\right)-\log_{10} R_{\alpha \beta}\right]
,\]
where $R_{\alpha \beta }$=2.86 in non-active galaxies, and $R_{\alpha \beta }$=3.10 in active galactic nuclei \citep{Ost89}.

We selected galaxies with $z<0.2$, $\Delta H_\alpha /H_\alpha <1/3$, $\Delta H_\beta /H_\beta <1/3$,
$0.1\le \overline{\gamma _V}(<1.5")\le 10$, $0.1<h_R({\rm kpc})<10,$ and
different ranges of absolute magnitude in r band ($M_r$; we did not correct for K--correction, which is small given the low redshift of the galaxies) between -23.0 and -18.0. Moreover, we required an error in the
value of the dust mass $M_d$ obtained through Eq. (\ref{md}) lower than 10\%. In total, we have 3551 galaxies with these constraints.

\begin{figure*}[htb]
\vspace{0cm}
\centering
\includegraphics[width=5.8cm]{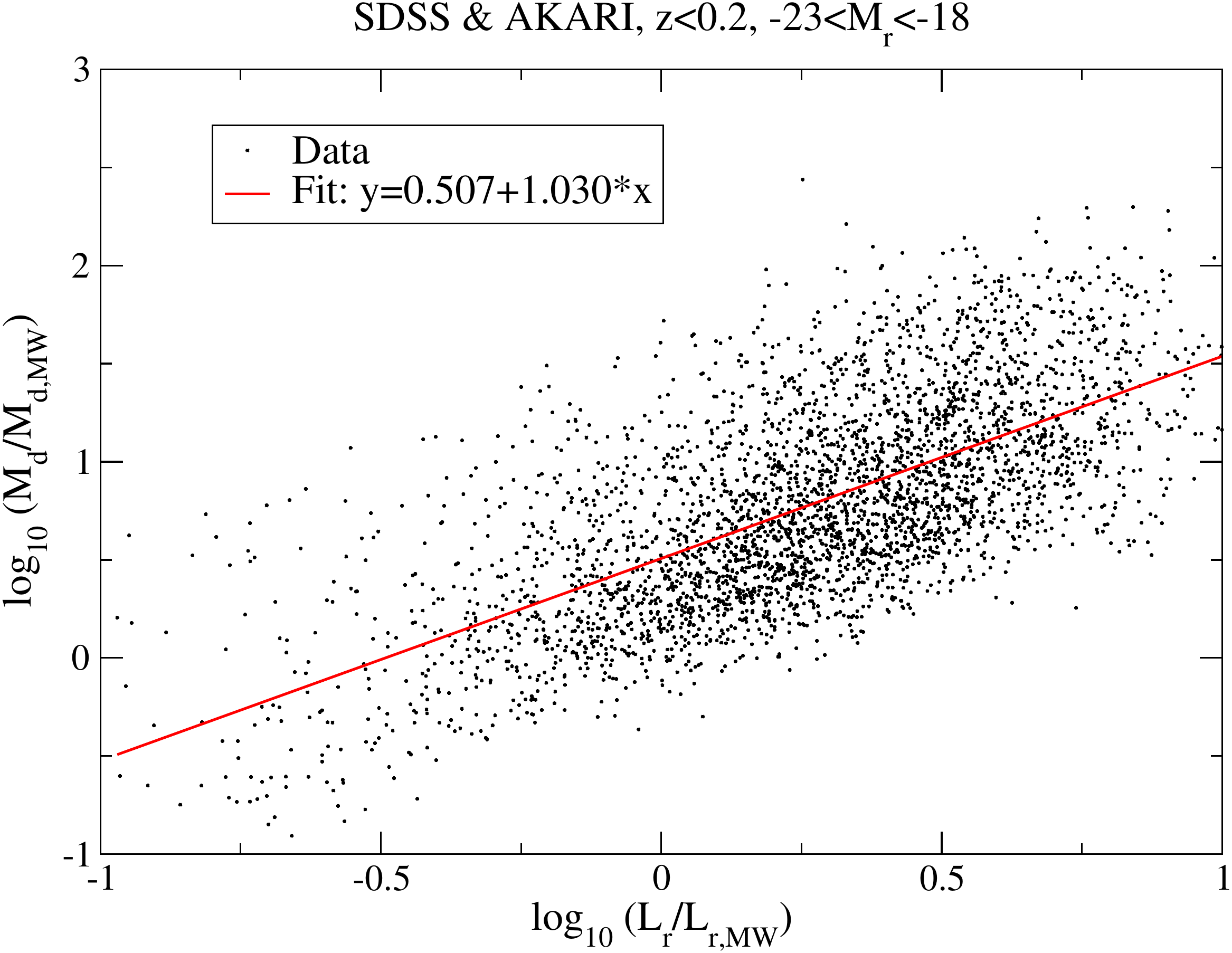}
\hspace{.2cm}
\includegraphics[width=5.8cm]{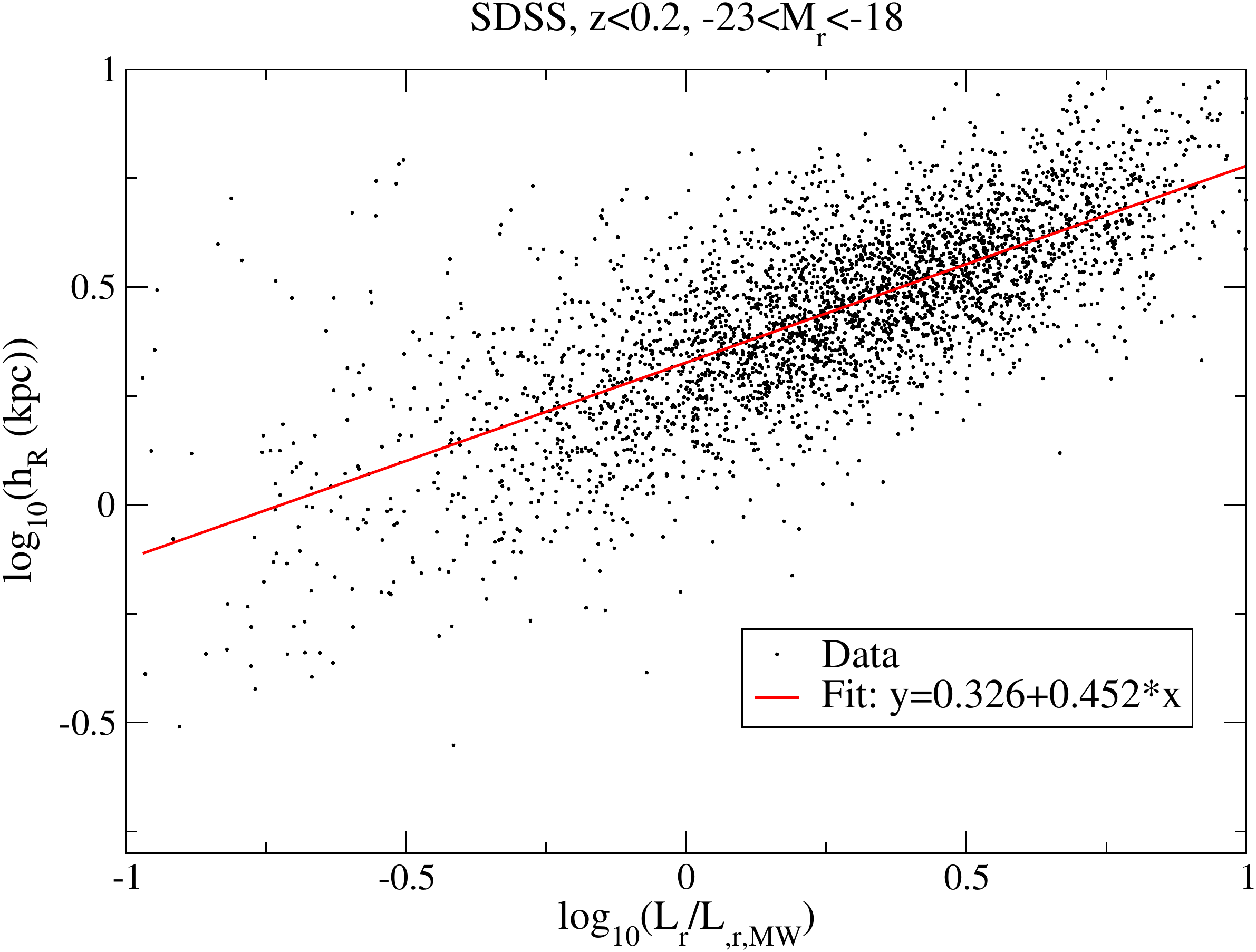}
\hspace{.2cm}
\includegraphics[width=5.8cm]{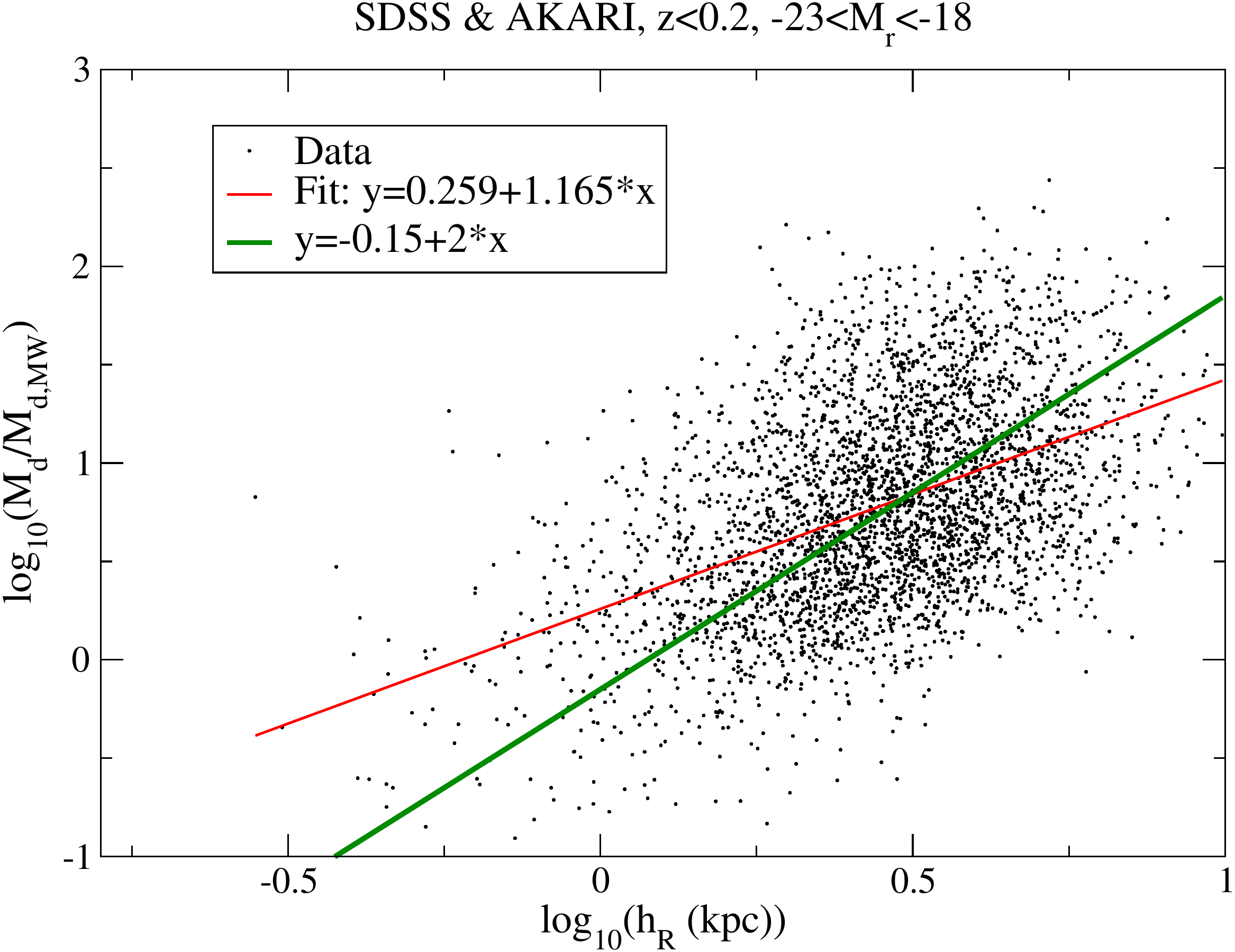}
\caption{Correlation of dust mass and optical luminosity (left), scalelength and optical luminosity (middle), and dust mass and scalelenth (right).}
\label{Fig:plots0}
\end{figure*}

The left panel of Fig. \ref{Fig:plots0} shows that the dust mass is approximately
proportional to the luminosity in r band. The central panel of Fig. \ref{Fig:plots0} shows that $h_R^2$ is approximately proportional to the luminosity in r band (we assumed $M_{r,MW}=-20.5$). 
This means through Eqs. (\ref{rhodisc}) and
(\ref{rhodustdisc}) 
that when a single exponential disk is assumed, the luminosity and dust surface densities at $R=0$ are approximately constant in the whole range of absolute magnitudes on average. 
The direct relation of the average dust mass with the scalelength is illustrated in the right panel of Fig. 
\ref{Fig:plots0}, giving a less clear trend. It forms a circle of points rather than an elongated structure, over which a linear trend is not clearly visible: a best linear fit in the log-log plot gives a slope 1.165, but as the green line shows, a slope of 2.00 is not excluded, therefore we can conclude
nothing from this last plot. Hence we assumed a relation $\langle M_d\rangle \propto \langle h_R\rangle ^2$
derived from the combination of the plots in the right and central panels of Fig. \ref{Fig:plots0}.

We carried out a double linear fit
of the type 
\begin{equation}
\label{fit1}
y=a+b*x_1+c*x_2,
\end{equation}\[
y=\log _{10}\left[\overline{\gamma _V}(<1.5")\left(\frac{M_d}{M_{d,MW}}\right)^{-1}\right],\]
\[x_1=\log_{10}h_R({\rm kpc}),\]
\[x_2=\log_{10}\left[\left(\frac{M_d}{M_{d,MW}}\right)\left(\frac{1}{h_R({\rm kpc})^2}\right)\right].\]
The reason for the choice of these variables is that 1) because the absorption is more or
less proportional to the total dust mass within the linear regime of low attenuation, we set
this variable $y$ which reflects the excess or deficit of absorption with respect to the amount of dust on it; 2) because of the relation found previously of $M_d\propto h_R^2$ on average (see Fig. \ref{Fig:plots0}), we set the independent variable $x_2$ as stated in order to
measure the effect of the variation in dust mass independently of the size
of galaxy. 
Rearranging the terms, this equation can be rewritten as
\begin{equation}
\label{fit2}
\log _{10}\overline{\gamma _V}(<1.5")=
a'+b'\log_{10}h_R({\rm kpc})
\end{equation}\[
+c'\log_{10}\left(\frac{M_d(h_R)}{M_{d,MW}}\right)
,\]
where $a'=a$, $b'=b-2c$, and $c'=1+c$.
However, Eq. (\ref{fit2}) cannot be
physically interpreted in a direct way because $M_d$ and $h_R$ are not independent variables, and
an increase in $h_R$ should be associated with an increase in $M_d$ on average.
Eq. (\ref{fit1})  physically represents the absorption per unit dust mass
as a function of scalelength and excess or deficit of dust mass for an average galaxy of a given
scalelength. We therefore relate the ratio of absorption to dust mass with two quantities that are almost independent.

\begin{figure}[htb]
\vspace{0cm}
\centering
\includegraphics[width=8.5cm]{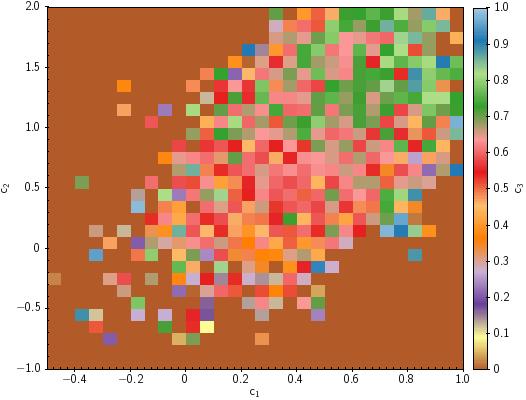}\\
\vspace{.2cm}
\includegraphics[width=8.5cm]{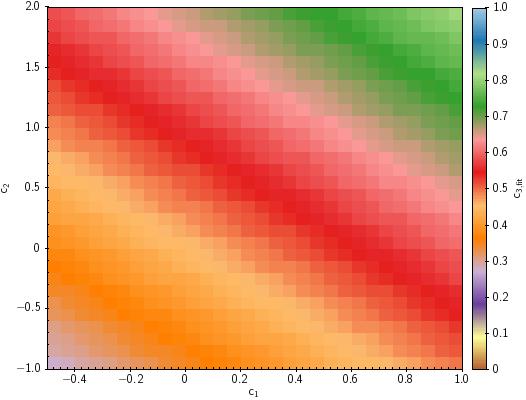}\\
\vspace{.2cm}
\includegraphics[width=8.5cm]{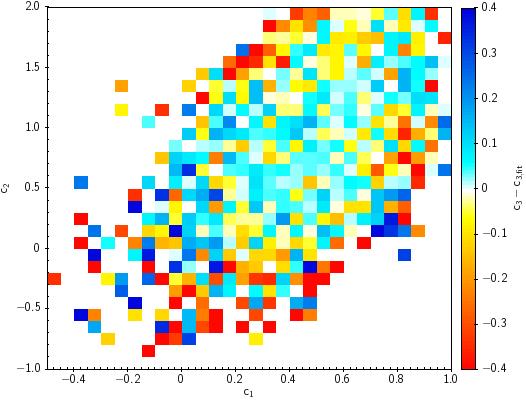}
\caption{Data (top panel; brown stands for an absence of data), best double-linear fit (middle), and residuals (bottom panel) of Eq. (\ref{fit2}): $c_3=\log _{10}\overline{\gamma _V}(<1.5")$ as a function of  
$c_1=\log_{10}h_R({\rm kpc})$ and $c_2\equiv \log_{10}\left(\frac{M_d(h_R)}{M_{d,MW}}\right)$.}
\label{Fig:dlinearfit}
\end{figure}

The result of the best fit gives $a=0.46$, $b=-1.63$, $c=-0.89$, with an r.m.s. of
$\sigma =0.26$, or equivalently, $a'=0.46$, $b'=0.15$, and $c'=0.11$.
In Fig. \ref{Fig:dlinearfit} we show the data, this best fit of $a'$, $b'$, and $c'$ 
, and the residuals. The dependence
is dominated by the choice of variables, which with a constant absorption would by default result in a $b'=0$, $c'=0$ dependence.

In these double-linear fits, an average departure of $\gamma _V\propto M_d$ is high, and the nonlinear
regime is evident. Our toy-model predictions in \S \ref{.toy} already foresaw this deviation for high values
of $M_d$, although it should be strictly linear for low values of $M_d$ (meaning $c=0$, $c'=1$). In order to test that the rough estimates of the model can represent the data approximately, we analyzed the dependence on the double-linear fit as a function of $M_d$. In Fig. \ref{Fig:depmdust} we plot
the dependence on $M_d$ when instead of taking the whole sample, different subsamples in different
ranges of dust mass in bins of 0.5 of $\log _{10} M_d/M_{d,MW}$ are used.
 $c'=1$ (absorption proportional to dust mass $M_d$) 
is obtained for low $M_d$, as expected, but the departure from linearity is conspicuous for $M_d>\sim M_{d,MW}$. For the remaining coefficients, $a'$ is more or less constant, and $b'$ presents a linear variation with dust mass, but within the r.m.s. of the fit (about 0.26). We therefore did not consider the variations of $b'$. 
The coefficient $c'$ 
follows the following dependence with $M_d$ from a fit of the data:
\begin{equation}
\label{cpMd}
c'=\exp\left[-(0.59^{+0.14}_{-0.10})\frac{M_d}{M_{d,MW}}\right]
.\end{equation}
The different samples have a different average redshift (see also Fig. \ref{Fig:depmdust}), which means
a different linear size aperture instead of the average 1.5 kpc. We assumed, as observed in 
the test with the toy model (\S \ref{.toy}; middle panel of Fig. \ref{Fig:gamma_hr}), that the shape of the
dependence on mass does not change with the average linear size aperture, however, only a constant would
change, to be added to the factor $a'$.

The above expressions also give the absorption only within 1.5" from the center of the galaxy, and in order
to estimate the average global absorption in the whole galaxy, we need to calculate 
$\overline{\gamma _V}(R_{0,max}=\infty)=C(1.5\ {\rm kpc})\,\overline{\gamma _V}(<1.5")$ where $C(1.5\ {\rm kpc})$ is model dependent. When we apply the values of $C$
given by Eq. (\ref{factorC}) to a purely exponential disk model, average the redshift equal to 0.05, and 
take the average dependence of $c'$ on $M_d$ given in Eq. ({\ref{cpMd}) into account, we obtain

\begin{equation}
\label{fit2i}
\log _{10}\overline{\gamma _V}(R_{0,max})=\infty)\approx
0.22-0.19\log_{10}h_R({\rm kpc})\ \ \ \ \  
\end{equation}\[
+\exp\left[-0.59\frac{M_d}{M_{d,MW}}\right]\,\log_{10}\left(\frac{M_d}{M_{d,MW}}\right); \ \ 
\sigma =0.27 
,\]
\begin{equation}
\label{fit1i}
\log _{10}\left[\overline{\gamma _V}(R_{0,max})=\infty)\left(\frac{M_d}{M_{d,MW}}\right)^{-1}\right]\approx
0.22 \ \ \ \ \ 
\end{equation}\[
+\left[2\exp\left[-0.59\frac{M_d}{M_{d,MW}}\right] -2.19\right]
\log_{10}h_R({\rm kpc}) 
\]\[
+\left( \exp\left[-0.59\frac{M_d}{M_{d,MW}}\right]-1\right)
\log_{10}\left[\left(\frac{M_d}{M_{d,MW}}\right)\left(\frac{1}{h_R({\rm kpc})^2}\right)\right]; 
 \ 
\sigma =0.27 
.\]

In Eq. (\ref{fit2i}), the last term is negative and monotonously increases with $M_d$, except for a 
small fluctuation around $M_d\sim 2.2M_{d,MW}$, but only with a negligible maximum amplitude of 0.09, which
is due the approximation of $c'$ to an exponential function; asymptotically, it converges to a value zero
for high dust masses. For $h_R>0.1$ kpc therefore $\overline{\gamma _V}<3.2$.

\subsection{Method 2 for calculating internal absorption}

Based on fitting of reddening in late-type galaxies, \citet{Cho09} reported an average relation 
of internal attenuation in r band and concentration index $c\equiv \frac{RP_{50}}{RP_{90}}$ and the absolute magnitude
\begin{equation}
\label{Cho09}
\gamma _r=1.06\left[-1.35(c-2.48)^2+1.14\right]\times
\end{equation}\[
\left[-0.223(M_r+20.8)^2+1.10\right]
,\]
applicable within $-21.95\le M_r<\le-18.65$, $1.74\le c\le 3.06$.
We assumed $\gamma _V=1.34\gamma _r$ \citep{Rie85}. This gives absorption values that are
constrained within $\gamma _V<1.78$.

Within the ranges of validity of Eq. (\ref{Cho09}), we have 3348 galaxies with $-21.95<M_r<-18.65$.
The dependences and considerations 
of Fig. \ref{Fig:plots0} are the same as with method 1 because they do not depend on the absorption.
Following the same type of analysis as with method 1, we arrive at the best double-linear fit
of $y=a+b*x_1+c*x_2$ with $y=\log _{10}\left[\overline{\gamma _V}(R_{0,max}=\infty)\left(\frac{M_d}{M_{d,MW}}\right)^{-1}\right]$ and $x_1=\log_{10}h_R({\rm kpc})$, 
$x_2=\log_{10}\left[\left(\frac{M_d}{M_{d,MW}}\right)\left(\frac{1}{h_R({\rm kpc})^2}\right)\right]$:
$a=0.10$, $b=-1.84$, $c=-0.96$, and an r.m.s. of $\sigma =0.10$; or equivalently, $a'=0.10$, $b'=0.08$, $c'=0.04$ for a fit of the type $y=a'+b'*x_1+c'*x_2$ with $y=\log _{10}\overline{\gamma _V}(R_{0,max}=\infty)]$
$x_1=\log_{10}h_R({\rm kpc})$, and $x_2=\log_{10}\left(\frac{M_d(h_R)}{M_{d,MW}}\right)$.

When we compare these numbers with those of method 1, we see that they are similar. However, the data of absorption
for individual galaxies calculated with both methods, although they presents an average correlation, do not agree well for individual galaxies:
see Fig. \ref{Fig:gamma_gamma} for 272 galaxies with $0.045<z<0.055$ [this is in the redshift range in which the
correction factor $C$ of Eq. (\ref{factorC}) is applied], with a best power-law fit of
$C(1.5\ {\rm kpc})\,\overline{\gamma _{V,method\ 1}} (<1.5")=$ 
$(1.58\pm 0.13) \overline{\gamma _{V, method\ 2}}(R_{0,max}=\infty)^{0.54\pm 0.19}$.

The dependence of $c'$ on the dust mass is also similar to that of method 1: 
$c'=\exp\left[-(1.47^{+0.19}_{-0.16})\frac{M_d}{M_{d,MW}}\right]$. 
In spite of the small discrepancies, the average dependence of the absorption
on size and dust mass is approximately the same as in method 1. The coefficients are summarized
in Table \ref{Tab:abc},\footnote{In this table, the third column ($N$) indicates the number of galaxies in each bin, the fourth column indicates the
average redshift, and the fifth column indicates the r.m.s. of the fit.} with the average of both methods. This average value of $b'=-0.05\pm 0.19$
agrees with the value of zero deduced with the theoretical model 
(see the left panel of Fig. \ref{Fig:gamma_hr}).

\begin{figure}[htb]
\vspace{0cm}
\centering
\includegraphics[width=8cm]{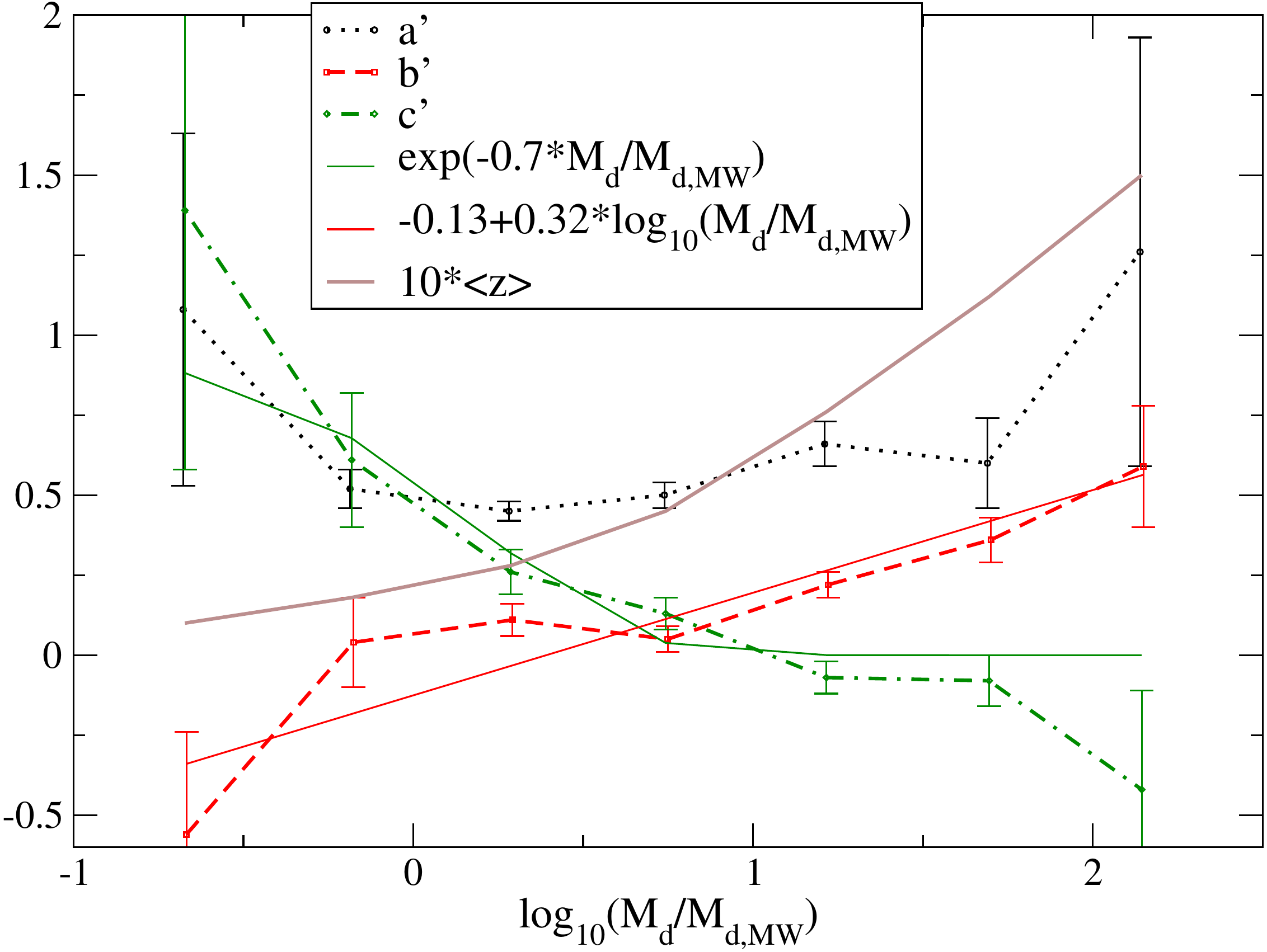}
\caption{Dependence on $M_d$ of the fit coefficients $\log _{10}\gamma _V=a'+b'\log_{10}h_R({\rm kpc})+
c'\log_{10}(M_d/M_{d,MW})$.}
\label{Fig:depmdust}
\end{figure}

\begin{figure}[htb]
\vspace{0cm}
\centering
\includegraphics[width=8cm]{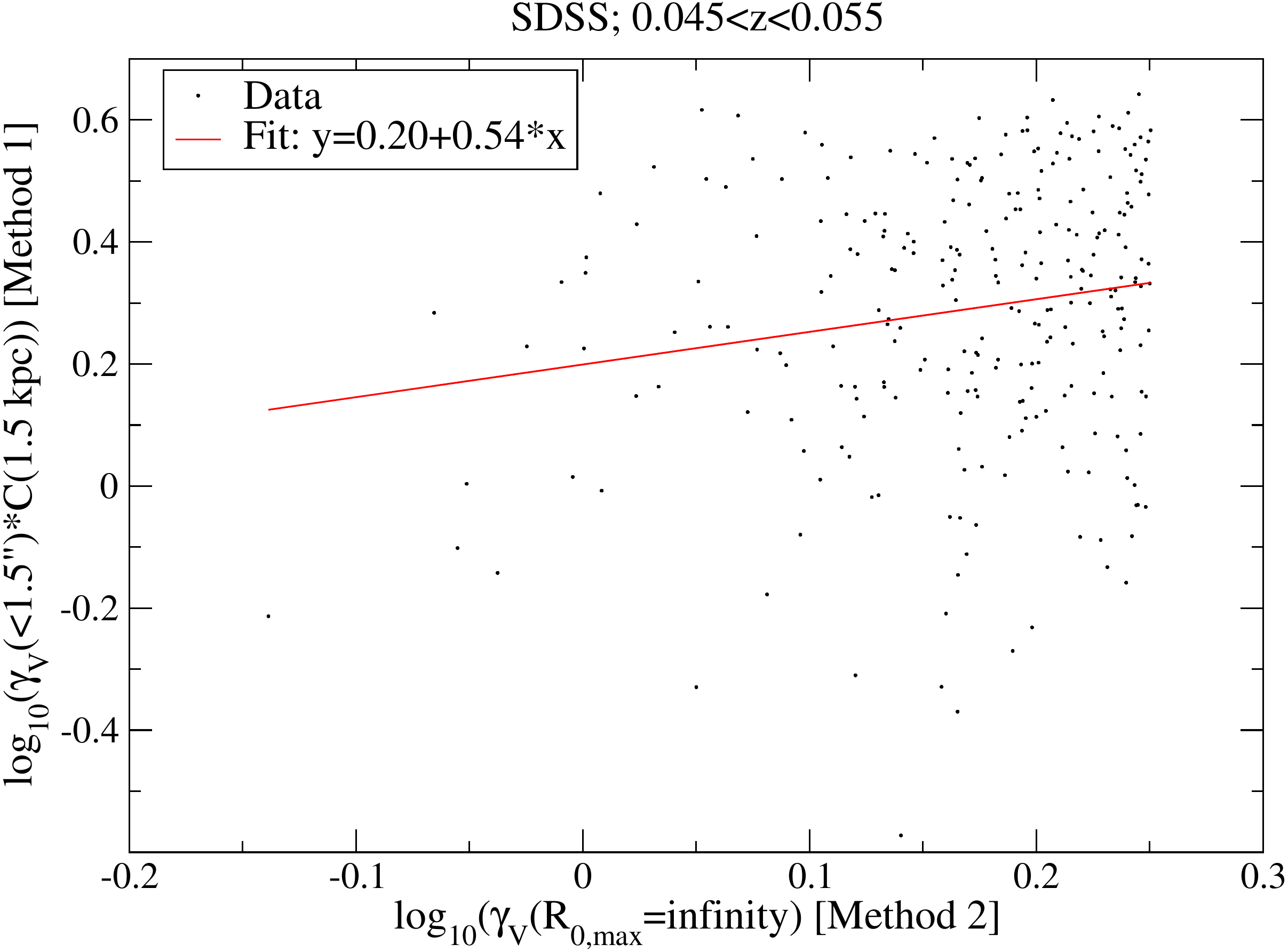}
\caption{Log-log plot of the correlation of the average absorption in the whole galaxy calculation with
method 1 (and the model-dependent correction of aperture $C$) and method 2.}
\label{Fig:gamma_gamma}
\end{figure}

\begin{table*}
\caption{Coefficients of the double linear fit $\log _{10}\gamma _V (R_{0.max}=\infty)=a'+b'*\log _{10}h_R({\rm kpc}) +c'*\log _{10}(M_d/M_{d,MW})$ for different ranges of dust mass $M_d$, with $\gamma _V$ calculated
with two different methods.} 
\begin{center}
\begin{tabular}{cccccccc}
Method & $\log _{10}(M_d/M_{d,MW})$ & $N$ \ & $\langle z\rangle $ & $\sigma $
& $a'$ & $b'$ & $c'$   \\ \hline \\
1 & -1.0\ \ ---\ \ 2.5 & 3551 & 0.054 & 0.27 & 0.22 & -0.19 & $\exp\left[-0.59\frac{M_d}{0.8M_{d,MW}}\right]$ \\ \\
2 & -1.0\ \ ---\ \ 2.5 & 3348 & 0.048 & 0.10 & 0.10 & 0.08  &  $\exp\left[-1.47\frac{M_d}{M_{d,MW}}\right]$ \\ \hline
Average & -1.0\ \ ---\ \ 2.5 & --- & 0.051 & --- & $0.16\pm 0.08$ & $-0.05\pm 0.19$ &  
$\exp\left[-(1.0\pm 0.6)\frac{M_d}{M_{d,MW}}\right]$ \\ \hline
\label{Tab:abc}
\end{tabular}
\end{center}
\end{table*}

\subsection{Attenuation as a function of FIR flux}
\label{.extFIR}

\subsubsection{Dust mass as a function of 100 $\mu m$ flux}

We can express the attenuation as a function of the FIR flux through the corresponding relation
of luminosity and flux, making use of the luminosity distance in a given cosmology.
Bolometric luminosities are not always accessible in FIR. A total sampling in many 
frequencies would be required. However, it can be approximately obtained with only one filter at rest around
100 $\mu m$ at rest: When
a fixed temperature $T_{\rm dust}\sim 20-25$ K \citep{Gal12} is assumed, in agreement with the 23 K we obtain from
our fits, the maximum emission is precisely at $\lambda \approx 100\ \mu {\rm m}$. 
We can also assume that 
\begin{equation}
\nu L_{100\mu {\rm m}}\approx f_{100}\,L_{bol}
\label{l100}
,\end{equation}
where $L_{bol}$ 
is the bolometric luminosity of dust, $\nu =\frac{c}{\lambda }$, $\lambda =100\ \mu$m. 
The parameter is usually set to $f_{100}\approx 1$ because $\nu $ 
is approximately the width of the luminosity distribution ($\Delta \nu $).
However, we can derive a better estimate of the parameter using our sample: 4379 galaxies from the
AKARIxSDSS sample with errors of the two luminosities lower than 20\%.
Fig. \ref{Fig:bollum} shows the relation of the bollometric luminosity and the luminosity corresponding
to 100\ $\mu$m (with AKARI fluxes at 90 and 140 $\mu$m, we interpolated the value of 
the observed frequency $100(1+z)$ $\mu $m flux). As a result of the best fit, we obtain 
$f_{100}=0.65$ on average ($\chi^2$ fit assuming $\Delta L_\nu /L_\nu $ constant).

\begin{figure}[htb]
\vspace{0cm}
\centering
\includegraphics[width=8cm]{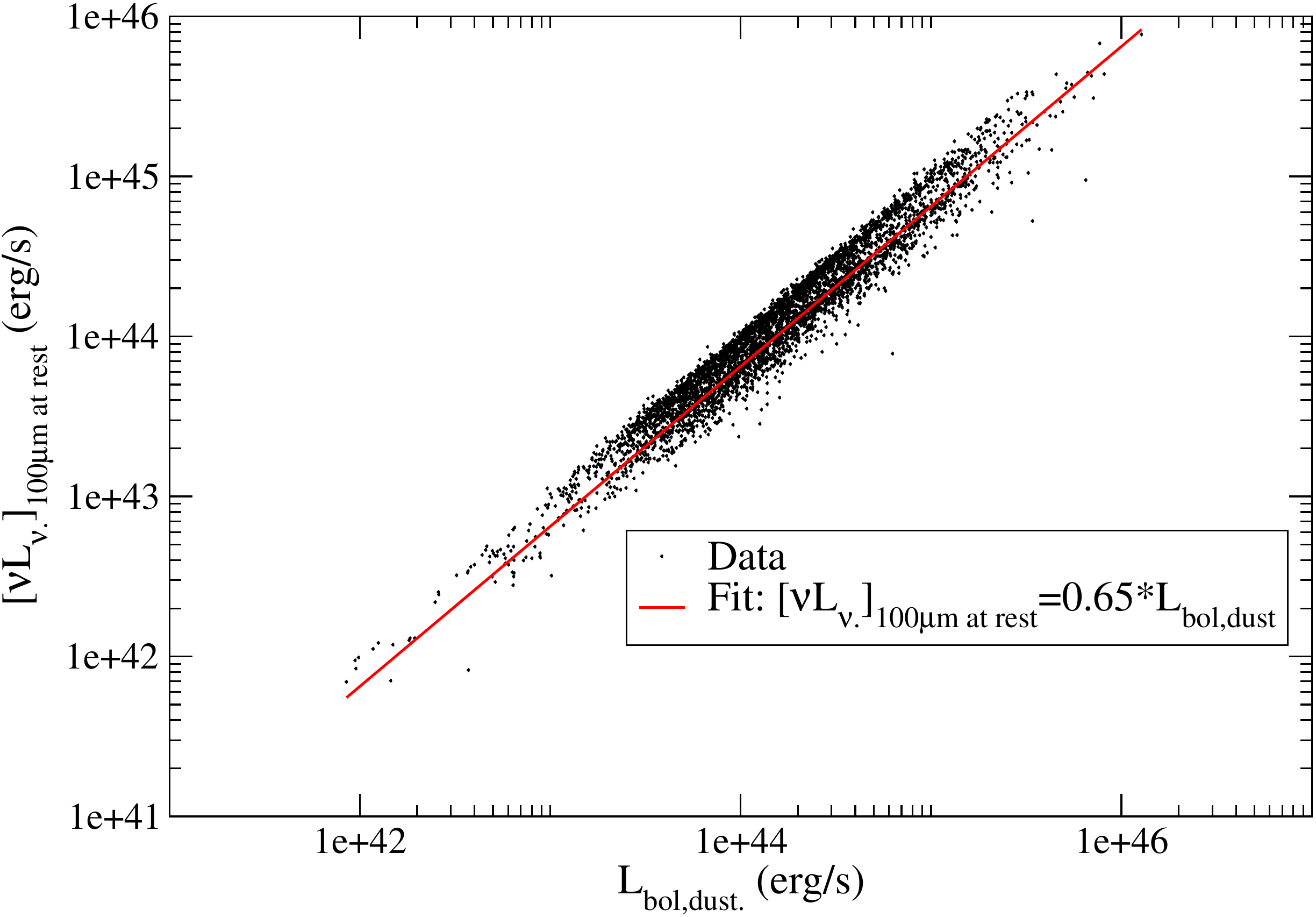}
\caption{Log-log plot of the correlation of the luminosities at 100$\mu $m (at rest) and bolometric luminosity for
4370 galaxies from the AKARIxSDSS sample with errors of the two luminosities lower than 20\%.}
\label{Fig:bollum}
\end{figure}

The FIR bolometric flux $F_{100\mu m(1+z)}$ is related to the luminosity $\nu L_{100\mu m(1+z)}$ by means of
\begin{equation}
F_{100\mu m(1+z)}(Jy)=
\frac{(1+z)\nu L_{100\mu m(1+z)}({\rm erg/s})}{4\pi d_L(z)({\rm Mpc})^2\times (3.00\times 10^{12}\ {\rm Hz})}
\end{equation}\[
\times \frac{1}{(10^{-23}\ {\rm erg/s/cm^2/Hz/Jy)}(3.086\times 10^{24}\ {\rm cm/Mpc})^2}
\]\[
=2.8\times 10^3\frac{(1+z)}{d_L(z)({\rm Mpc})^2}\frac{\nu L_{100\mu m(1+z)}} {10^{43}\ {\rm erg/s} }
,\]
where $d_L(z)$ is the luminosity distance.

When we relate the luminosity at 100 $\mu$m at rest with the dust mass through
Eq. (\ref{md}), with the bolometric luminosity related according to Eq. (\ref{l100}), $f_{100}=0.65$, 
taking the average values of $a'$ and $b'$ from Table \ref{Tab:abc}, and neglecting
the dependence with $h_R$ (because $b'=-0.05\pm 0.19$ is compatible with zero, which is the expected
theoretical value), we obtain that the attenuation is
\begin{equation}
\label{gvf}
\overline{\gamma _V}=(1.45\pm 0.27)f^{\exp{[-(1.0\pm 0.6)f}]} 
,\end{equation}\[
f=\frac{M_d}{M_{d,MW}}=\frac{F_{100\mu m(1+z)}}{700\ {\rm Jy}}
\frac{d_L(z)({\rm Mpc})^2}{(1+z)}
,\]
or equivalently, applying the definition of Eq. (\ref{gammaV}),
\begin{equation}
\label{AVlow}
\overline{A _V}=(1.45\pm 0.27)f^{\exp{[-(1.0\pm 0.6)f}]} \log _{10}\left(\frac{r_{\rm major}}{r_{\rm minor}}\right) 
.\end{equation}

As discussed in \S \ref{.toy}, this negligible dependence on the galaxy size is kept while
a relation $M_d\propto h_R^2$ is given for most of the galaxies. The data indicate that this is a good
approximation for low redshift. However, as we show in \S \ref{.highz}, the ratio
of dust mass and size may not be the same.

\subsubsection{Regimes of low and high dust mass}

Two special cases can be observed. For very high dust masses ($M_d\gtrsim 10^8$ M$_\odot$, $f\gtrsim 3,6$), 
we may consider the term $c'\approx 0$ (saturation of absorption),
so that the absorption is $\overline{\gamma_V}=(1.45\pm 0.27)$.
For very low dust masses ($M_d\lesssim 10^7$ M$_\odot$, $f\lesssim 0.36$), we may consider $c'\approx 1$ (linear relation
of absorption and dust mass), so that we can consider the following dependence: 
$\overline{\gamma _V}=(1.45\pm 0.27)f$ (with $f<<1$).

\section{Evolution of the ratio of dust absorption to emission}
\label{.highz}

The dependence of the absorption on the galaxy size and FIR luminosity (implicit in the dependence on the dust mass) is interesting because this might be used as a cosmological test of the galaxies at high $z$.
We did not carry out an analysis at high $z$ given the lack of data at present, but we provide
some indications how it might be used.

\subsection{General law for any $M_d(h_R)$}

The equation of the ratio of absorption to emission is interesting for studying high-redshift galaxies because it is not affected by evolution.
It is independent of the stellar luminosity evolution and only depends on the dust properties (parameters
$\beta $ and $T_d$), but we
know that the evolution of dust properties with $z$ in spiral galaxies is not significant \citep{Lop17b}.
Only the possible evolution in size is expected to affect this evolution. 

The stellar mass and dust mass are roughly proportional in spiral
galaxies \citep{Cal17,Dav19}, and spiral galaxies with the same stellar mass have sizes at high $z$ that are much smaller
than at low $z$ \citep{Tru06}. When we assume that this evolution of galaxy sizes is correct,
we therefore have much smaller $h_R$ at high $z$ than at low $z$ for the same dust masses.
The equations derived in \S \ref{.local} are not valid in the application to high $z$ accordingly because
the proportionality between dust mass and size changes. Nonetheless, from \S \ref{.local} we learned that our basic predictions of the toy model in \S \ref{.toy} are roughly correct: in particular,
that the dependence on the size is almost null when $M_d\propto h_R^2$ because the decrease in
absorption with increase in size is compensated for by the increase in dust mass. 
The question now is what happens in an hypothetical case in which the size increases or decreases, but the
dust mass is kept constant. In this case, the dependence on size is not null because the dust mass does not vary in the way predicted in the left panel of
Fig. \ref{Fig:gamma_hr}. 

At low $z$, we obtain 
\begin{equation}
\label{mdhrlow}
M_d=M_{d,MW}\times 10^{-0.15}\times h_R({\rm kpc})^2
\end{equation} 
(see the right panel of Fig. \ref{Fig:plots0}). A null dependence on size can be 
understood as a factor $f$ in Eq. (\ref{gvf}),
\begin{equation}
f(h_R)=10^{0.075\beta}\times h_R({\rm kpc})^{-\beta}\times \left(\frac{M_d}{M_{d,MW}}\right)^{(1+0.5\beta )}
.\end{equation}
When we use $M_d$ from Eq. (\ref{mdhrlow}), we recover $f(h_R)=\left(\frac{M_d}{M_{d,MW}}\right)$. However, when high $z$ is compared with low
$z$,  $M_d$ is kept constant and $h_R$ changes at high $z$ with respect to the low $z$ cases,
an inconstant $f(h_R) $ would result. In a general case, taking $\beta =1.75$ as obtained in the right panel of Fig. \ref{Fig:gamma_hr} in the linear regime, Eq. (\ref{gvf}) would be expressed as 
\begin{equation}
\label{gvf2}
 \overline{\gamma _V}=(1.45\pm 0.27)f_M^{\exp{[-(1.0\pm 0.6)f_M]}} 
,\end{equation}\[
f_M=1.35\times h_R({\rm kpc})^{-1.75}\times \left(\frac{M_d}{M_{d,MW}}\right)^{1.87} 
\]\[
=7.6\times 10^{-6} \alpha_{hR}^{-1.75}\times
\left(\frac{F_{100\mu m(1+z)}}{700\ {\rm Jy}}\right)^{1.87}
\times f_{\rm cosmol.}(z) 
,\]\[
f_{\rm cosmol.}(z)=\left(\frac{d_L({\rm Mpc})^{3.75}}{(1+z)^{1.87}d_A({\rm Mpc})^{1.75}}\right)
,\]
where $\alpha _{hR}$ is the equivalent angular size of the scalelength $h_R$, and $d_A$ is the 
angular distance. With $f_M$ we represent $f$ for a fixed value of $M_d$.
Only the factor $f_{\rm cosmol.}(z)$ depends on
 the cosmological model. For any cosmological model, $d_A=(1+z)^{-\eta} d_L(z)$ [in the case of
a Friedmann-Lema\^itre-Robertson-Walker cosmology, $\eta =2$, what is called Etherington's distance-duality
relation \citep{Hol11}], therefore
\begin{equation}
f_{\rm cosmol.}(z)=d_L(z)({\rm Mpc})^2(1+z)^{(1.75\eta -1.87)}
.\end{equation}
Figure \ref{Fig:cosmologies} shows the dependence of $f_{\rm cosmol.}$ on the
cosmological models,\footnote{The 
nine cosmological models are given by \citet[\S 2]{Lop16}. Note that all of the cosmological models follow Etherington's relation
($\eta =2$) except the static models, which have $\eta $ equal to 0.5, 0.5, 1.5 respectively for linear Hubble
law, single tired light and plasma tired light.}  thus giving the possibility of using the above relation as a 
cosmological test. The differences are especially large for static cosmologies that
do not follow Etherington's relation, therefore the test would be very suitable for testing the
value of $\eta $, even at $z<1$.

\begin{figure}[htb]
\vspace{0cm}
\centering
\includegraphics[width=8cm]{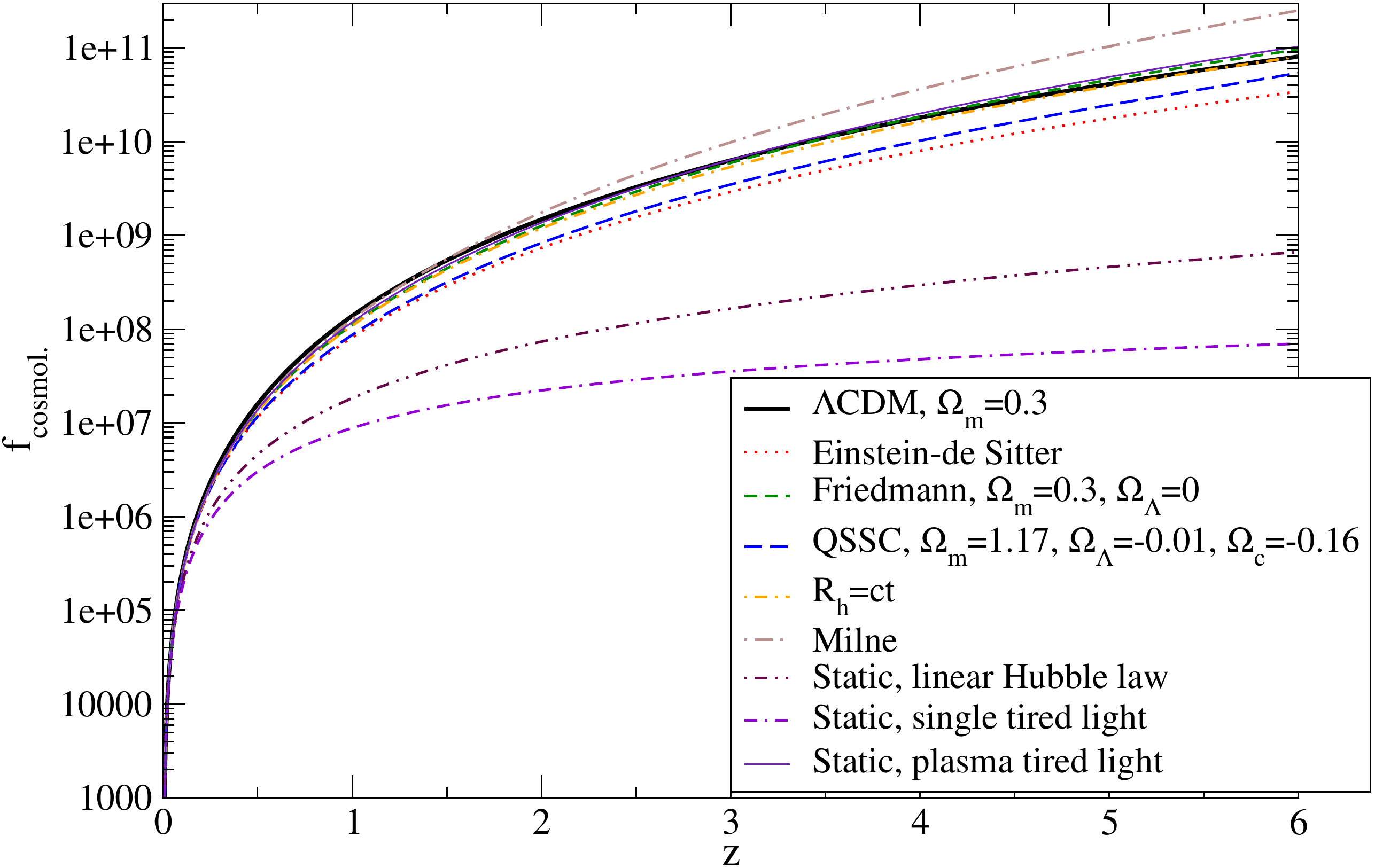}
\caption{Predictions of $f_{\rm cosmol.}(z)$ for nine cosmological
 models.}
\label{Fig:cosmologies}
\end{figure}

Equation (\ref{gvf2}) should not be interpreted as a direct dependence in the linear regime
of the absorption proportional to $M_d^{1.87}$; this is not the physical meaning because $M_d$ is not independent of $h_R$. 
It is a simple rule
to compare low- and high-redshift data, which means that when we fix the value of the dust mass, we can
evaluate how a change in galaxy size with respect to the low $z$ cases increases the
absorption.

\subsection{Difficulties in applying Eq. (\ref{gvf2}) to data at high $z$}

In order to use this expression in practical cases and show the variation of
absorption with size for a given dust mass, we need to be in the regime of nonsaturation, that is,
$f\lesssim 3.6$. In Fig. \ref{Fig:sens} we plot the required sensitivity of the FIR photometer assuming the
standard cosmological model.
As observed, we would need a detector that reached well below 0.5 mJy at $z>1$. At $z=3$, we would need
to detect sources with $F_{400\mu m}$ below 10-100 $\mu $Jy.

In Table \ref{Tab:surveysFIR} we list available FIR and millimeter surveys. No survey reaches the required limit of nonsaturation for high $z$. The only survey that approaches the limits  is ALMA at 1 mm, which can reach 47.5 $\mu $Jy, although
we would only observe the peak of dust emission for sources at $z\approx 9$ at these wavelengths, which requires still
better sensitivity, and for lower $z,$ we expect lower fluxes at 1 mm, which would again require better
sensitivities. With current surveys, we therefore cannot analyze the separate
dependence on radius and mass because we would be in the saturation regime.
Nonetheless, it is only a matter of time that some FIR or millimeter survey with the
required sensitivity become available, or some explicit very deep observations could be carried out in some few 
high $z$ galaxies in order to reach the limits. Because of the strong dependence of $f_{{\rm cosmol.}}$ on
the cosmological model, current surveys might moreover obtain nonsaturated dust sources at high $z$ for some exotic
cosmological models (see Fig. \ref{Fig:cosmologies} or \S \ref{.example}).

A greater difficulty might be obtaining a good absorption measurement. The application of the
method
of the line ratios of H$_\alpha $ and H$_\beta $ (method 1) is possible and was already obtained
for some high $z$ galaxies \citep[e.g.,][]{Mas14,Sch18} (note that these lines are observed in the near-infrared
at high $z$).
However, we obtain a dispersion of 
$\log _{10}\overline{\gamma _v}$ equal to 0.28 [Eq. (\ref{fit2i})], and at higher $z,$ the dispersion might
be higher because noisier spectra are used. Method 2 cannot be applied directly to
high $z$ because its calibration was only valid at low $z$ and depends precisely on the size evolution
that we wish to test. Another method, method 3, might be used to derive the absorptions. It uses photometry in many filters and fits the SED with galactic templates, where the 
mean absorption $\overline {A_V}$ is a free parameter,
as was already obtained for $z>2$ galaxies \citep[e.g.,][]{Ono10,Gla17,Sch18}.
The absorption can also be estimated using the reddening, that is, the
excess of intrinsic colors: for instance, \cite{Cho09} (the basis of our method 2)
have used at low $z$ the colors $r-K$, $u-K$ or $u-r$ to statistically estimate the
absorption of galaxies. Regardless of the method used to derive the absorption, our guess
is that the errors of individual sources will be large,
and we will need too many galaxies at high $z$ to explore the relation successfully.

The last quantity that is necessary to carry out a galaxy size test at high $z$ is the galaxy size using
different cosmological models (\citet{Tru06,Shi15} for the standard cosmological model
or \citet{Lop10} for different cosmological models). When we have the three amounts, absorption, dust
mass derived from FIR emission, and galaxy size for a given cosmology, we can verify that
they follow Eq. (\ref{gvf2}), corroborating or rejecting the cosmological model that was used.

\begin{figure}[htb]
\vspace{0cm}
\centering
\includegraphics[width=8cm]{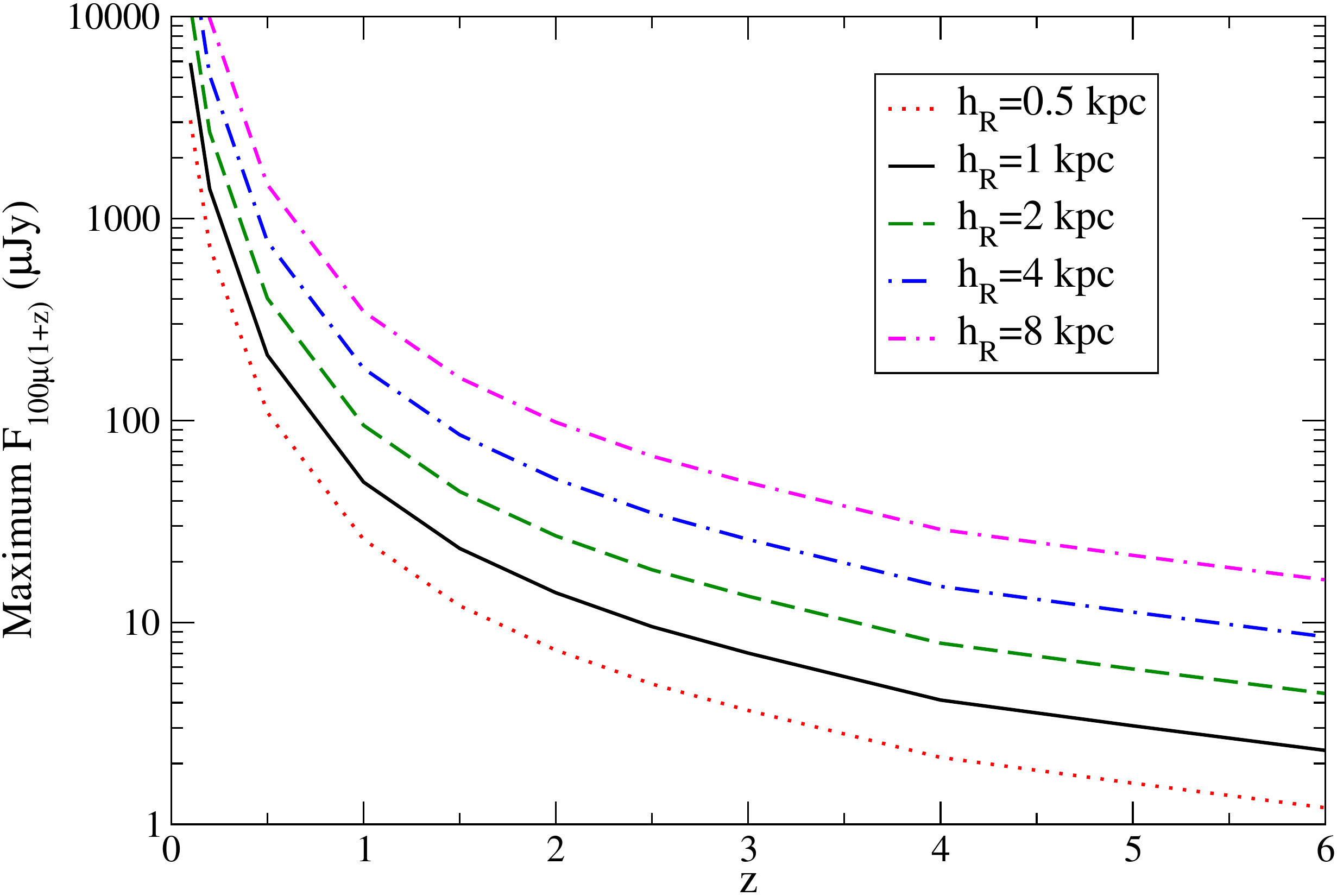}
\caption{Minimum sensitivity of a significant detection of a photometer at a wavelength $(1+z)100\ \mu m$ 
as a function of the redshift and the scalelength of the galaxy, derived making $f_M=3.6$ in
Eq. (\ref{gvf2}) using standard cosmological $\Omega _\Lambda =0.7$, $H_0=70$ km/s/Mpc
to evaluate the luminosity distance.}
\label{Fig:sens}
\end{figure}

\begin{table*}
\caption{FIR and millimetric wavelength $\le 1.5$ mm surveys and their sensitivity at 5$\sigma $ detection.}
\begin{center}
\begin{tabular}{ccccc}
Survey, instrument & Fields & Wavelenth ($\mu $m) & Minimum flux ($5\sigma $) ($\mu $Jy) & Reference \\ \hline
Spitzer, MIPS & EGS, GOODS-N/-S, ECDFS & 70 & 2500-3500 & \citet{Mag09} \\
Herschel/SPIRE, H-ATLAS & 500 sq. degrees & 250 & 25500-60000 & \citet{Eal10} \\
Herschel/SPIRE, H-ATLAS & 500 sq. degrees & 350 & 28000-60500 & \citet{Eal10} \\
Herschel/SPIRE, H-ATLAS & 500 sq. degrees & 500 & 33000-65000 & \citet{Eal10} \\
Herschel/SPIRE, HerMES & several & 250 & 3800-53000 & \citet{Oli12} \\
Herschel/SPIRE, HerMES & several & 350 & 3100-53000 & \citet{Oli12} \\
Herschel/SPIRE, HerMES & several & 500 & 4500-76500 & \citet{Oli12} \\
Herschel/PACS, PEP & several & 70 & 1700 & \citet{Lut11} \\
Herschel/PACS, PEP & several & 100 & 900-7500 & \citet{Lut11} \\
Herschel/PACS, PEP & several & 170 & 2200-16400 & \citet{Lut11} \\
ALMA, ASPECS & HUDF & 1000 & 63.5 & \citet{Wal16} \\
ALMA, ASPECS LP & HUDF & 1000 & 47.5 & \citet{Wal16} \\
ALMA, GOODS & GOOD-S, 70 sq. arcmin. & 1100 & 900 & \citet{Fra18} \\
ALMA, ASAGAO & GOOD-S, 26 sq. arcmin. & 1200 & 190 & \citet{Hat18} \\ \hline
\label{Tab:surveysFIR}
\end{tabular}
\end{center}
\end{table*}

\subsection{Example of an application of Eq. (\ref{gvf2})}
\label{.example}

We illustrate how Eq. (\ref{gvf2}) should be applied for the cases in the future when we have enough data, or for cases
of alternative cosmologies that give a $f_{{\rm cosmol.}}$ much lower than $\Lambda $CDM cosmology.
We assumed that we have a spiral galaxy at $z=4$  and inclination $i=60\pm 5$
 deg.
We also assumed that we measured an angular scalelength (or the Petrosian angular size, which
can later be converted into the scalelength through Eq. (\ref{scale_petro})) of $\alpha_{hR}=0.20\pm 
0.02$ arcsec, and we measured a flux $F_{500 \mu m}=(100
\pm 20)$ $\mu $Jy.
Dust mass and sizes may be calculated depending on the cosmological 
model. For illustration,
we give two very different examples (whether they are realistic or not is not the 
discussion here).

\begin{itemize}
\item Using the standard $\Lambda $CDM cosmological model with $H_0=69.6$ km/s/Mpc, $\Omega _m=0.286$, 
at $z=4$
the luminosity distance $d_L$ and the angular
distance $d_A$ are $d_L=d_A\,(1+z)^2=3.65\times 10^4$ Mpc; $f_{{\rm cosmol.}}=1.81\times 10^{10}$. The linear size scalelentgh is $h_R=1.41\pm 0.14$ kpc.
With the relation of flux and dust mass given in Eq. (\ref{gvf}), this means  $M_d=(10.5\pm 2.2)\times 10^8$
M$_\odot $. When we now apply Eq. (\ref{gvf2}), we obtain $f=113\pm 23$, which is hypersaturated because $f$ is much
larger than 3.6: $\overline {\gamma _V}=1.45\pm 0.27$,
or using Eq. (\ref{gammaV}), $\overline {A_V}=0.44\pm 0.12$ magnitudes.

\item For a simple static Euclidean universe with redshift due to a simple tired light effect
\citep{Lop10,Lop16}, with $H_0=69.6$ km/s/Mpc, at $z=4$
the luminosity distance $d_L$ and the angular
distance $d_A$ are $d_L=d_A\,(1+z)^{1/2}=1.55\times 10^4$ Mpc; $f_{{\rm cosmol.}}=4.81\times 10^7$. The linear size scalelength is $h_R=6.7\pm 0.7$ kpc.
With the relation of flux and dust mass given in Eq. (\ref{gvf}), this means  $M_d=(1.88\pm 0.38)\times 10^8$
M$_\odot $. When we now apply Eq. (\ref{gvf2}), we obtain $f=(0.31\pm 0.07)$, which in the linear regime is 
$\overline {\gamma _V}=0.45\pm 0.13$,
or using Eq. (\ref{gammaV}), $\overline {A_V}=0.14\pm 0.05$ magnitudes.

\end{itemize}

We are probably unable to measure $\overline {A_V}$ with enough precision to distinguish between the two
results, but when the analysis is performed with many galaxies, it will be possible to statistically distinguish the two
cosmologies. We have used an example that saturates in the standard cosmological model at $z=4$. If
we were able to measure much lower FIR fluxes, below 10 $\mu $Jy, the case would become even more interesting because
the difference in absorption between the two models is distinguished by a factor larger than 100.
At lower $z$, for instance, $z\lesssim 1,$ the distinction between the different cosmologies is small,
but the sensitivity of FIR photometers might be enough with our present technology.

\section{Conclusions} 

We have derived an average relation of the internal mean absorption of
galaxies in $V$ filter (for another filter, we just need to multiply it by some factor \citep[e.g.,][]{Rie85})
and the dust mass derived from FIR emission at low $z$: Eq. (\ref{AVlow}), finding that the dependence
on galaxy size is negligible. A smaller galaxy for the same amount of dust would give
larger absorption, but because $M_d\propto h_R^2$ on average, this dependence on size cancels out.
This equation allowed us to derive the internal attenuation of the galaxy only knowing its 100 $\mu $m flux.

There is a maximum value of $\gamma_V \equiv \frac{\overline {A_V}}{\log _{10}\left(\frac{1}{\cos i}\right)}$
of $1.45\pm 0.27$ magnitudes because when a dust density is exceeded, the absorption saturates because
we cannot see the stars in the background of a thick layer of dust and we only see the last scattering
surface of the galaxy. This saturation limit is reached approximately for dust masses higher than $10^8$
M$_\odot $ for low $z$. For dust mass lower than $10^7$ M$_\odot $ , we are in a linear regime, and
 the mean absorption is proportional to the dust mass.

When this law is extrapolated to high $z$ galaxies, we must bear in mind that the relation of  $M_d\propto h_R^2$ observed at low $z$ is not kept because for a constant dust mass, we have much smaller sizes due
to the putative galaxy size evolution necessary to make the data compatible with standard cosmology. Considering the observational results of low $z$ galaxies and the toy model
we developed, we can derive a equation that relates the internal mean absorption of
galaxies, the dust mass derived from FIR emission at low $z,$ and the scalelength of the galactic disk
(or the Petrosian radius). This is given in Eq. (\ref{gvf2}). 
This might be used as a cosmological test because the factor $f_{{\rm cosmol.}}$ (to which 
the absorption is proportional in the linear regime) at high $z$ 
varies strongly in different cosmologies.
Although the capabilities of the present-day FIR and millimeter surveys do not allow us to carry out this
cosmological test at present within the standard model, it may be used in the future, when, for instance, we can observe  galaxies at $z=3-5$ with a sensitivity better than $\sim 10\ \mu$Jy. For alternative very different
models such as a static Universe, the application of the Eq. (\ref{gvf2}) would predict very low absorptions
in detectable FIR fluxes at high $z$.

\begin{acknowledgements}
Thanks are given to Helmut Dannerbauer for providing information of Table \ref{Tab:surveysFIR}. 
Thanks are given to the anonymous referee for very useful comments and suggestions that helped
to improve this paper. 
Thanks are given to the language editor of A\&A Astrid Peter for proofreading of this text.
M.L.-C. was supported by the grant PGC-2018-102249-B-100 of
the Spanish Ministry of Economy and Competitiveness (MINECO).
Based on observations with AKARI, a JAXA project with the participation of ESA.
Funding for the SDSS and SDSS-II has been provided by the Alfred P. Sloan Foundation, the Participating Institutions, the National Science Foundation, the U.S. Department of Energy, the National Aeronautics and Space Administration, the Japanese Monbukagakusho, the Max Planck Society, and the Higher Education Funding Council for England. The SDSS Web Site is http://www.sdss.org/.
The SDSS is managed by the Astrophysical Research Consortium for the Participating Institutions. The Participating Institutions are the American Museum of Natural History, Astrophysical Institute Potsdam, University of Basel, University of Cambridge, Case Western Reserve University, University of Chicago, Drexel University, Fermilab, the Institute for Advanced Study, the Japan Participation Group, Johns Hopkins University, the Joint Institute for Nuclear Astrophysics, the Kavli Institute for Particle Astrophysics and Cosmology, the Korean Scientist Group, the Chinese Academy of Sciences (LAMOST), Los Alamos National Laboratory, the Max-Planck-Institute for Astronomy (MPIA), the Max-Planck-Institute for Astrophysics (MPA), New Mexico State University, Ohio State University, University of Pittsburgh, University of Portsmouth, Princeton University, the United States Naval Observatory, and the University of Washington.

\end{acknowledgements}

\end{document}